# Rank ordering of species sensitivities to chemicals: discrepancies between behavioural and lethal responses in freshwater and marine invertebrates

Julien Mocq [a*], George Ruck [a b], Michaël Demortier [a], Flavie Brun [a], Romain Coulaud [c], Olivier Geffard [a] and Arnaud Chaumot [a*]

[a] INRAE, UR RiverLy, Laboratoire d'écotoxicologie, F-69625 Villeurbanne, France
[b] Viewpoint, 67 Rue Copernic 01390 Civrieux, France
[c] Université Le Havre Normandie, Normandie Univ, FR CNRS 3730 SCALE, UMR-I 02 SEBIO, F-76600, Le Havre, France
* Emails: julien.mocq@gmail.com ; arnaud.chaumot@inrae.fr

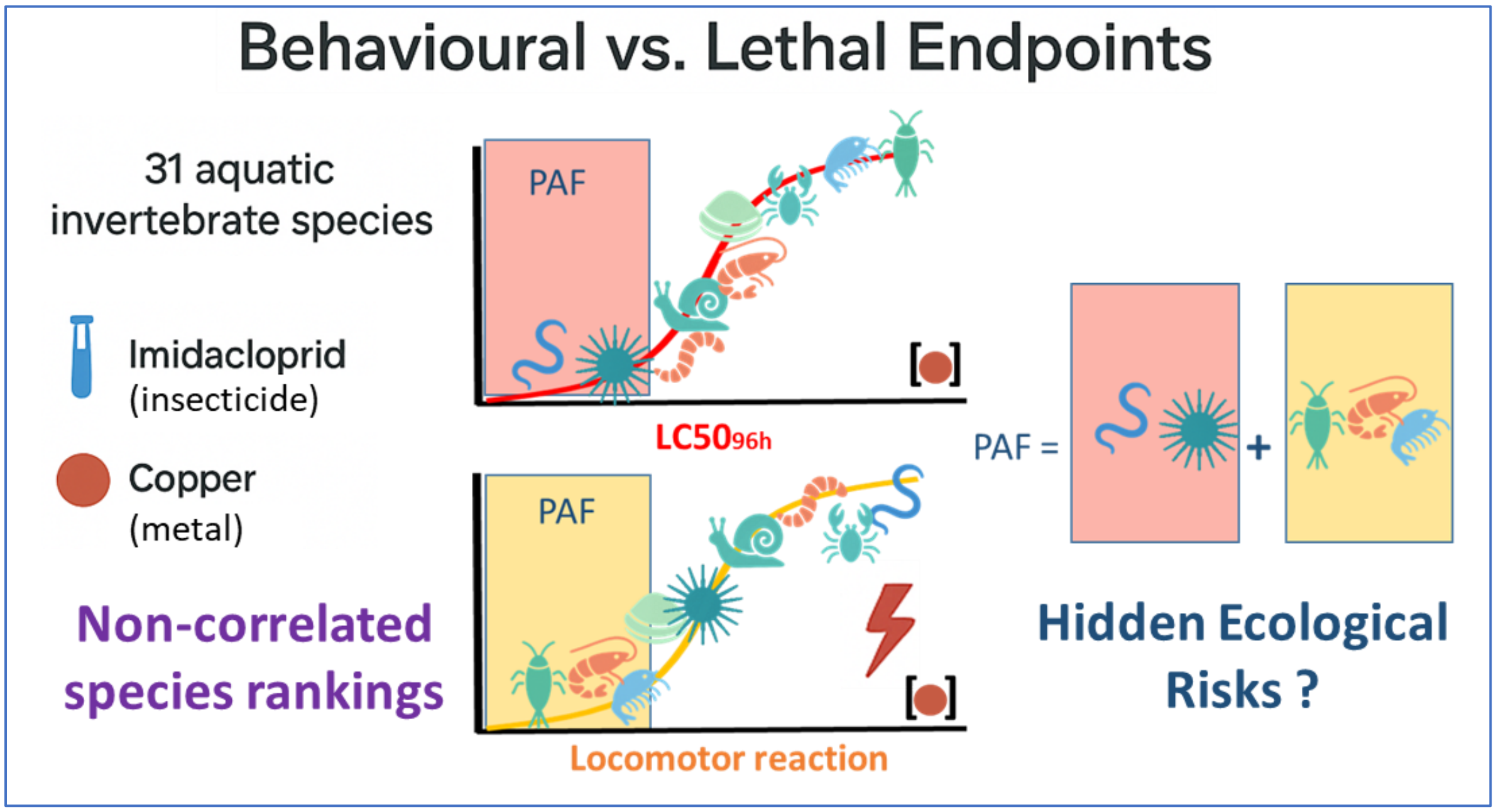

# 1. Abstract

Locomotor behavioural responses are increasingly recognized as sensitive and ecologically relevant indicators for assessing aquatic organism exposure to contaminants. They often occur at concentrations lower than those causing physiological damage, yet their consequences for organisms' fitness and population dynamics remain unclear. Moreover, it is still uncertain how these behavioural endpoints align with traditional toxicological apical endpoints used to assess species sensitivities in risk assessment. This study evaluates the correlation between thresholds triggering locomotor reactions with LC50 values across 19 freshwater and 12 marine invertebrate species exposed to the insecticide imidacloprid and the metal copper. Behavioural responses were measured via videotracking during pulse exposure assays, while 96h acute tests provided species-specific LC50 values for copper. Results reveal strong discrepancies: some highly sensitive species showed no locomotor changes, whereas others tolerant to lethality displayed strong movement responses. For instance, freshwater annelids and gastropods responded markedly to imidacloprid, while marine shrimps exhibited hyperactivity under low copper levels. These findings indicate that relying solely on lethality-based data may underestimate the fraction of species affected by contamination. Integrating behavioural endpoints into population risk assessment frameworks is therefore crucial to better protect aquatic biodiversity.

**Synopsis :** Survival tests are main indicators of species sensitivity in chemical risk assessment. This study reveals a total divergence with short-term behavioural locomotor reactions in aquatic invertebrates, questioning current acute risk assessment.

## 2. Introduction

Although traditionally less studied than mortality, growth or reproductive impairments, the use of behavioural response to assess chemical hazard spreads rapidly in recent aquatic ecotoxicology.[1,2] Behavioural endpoints include locomotion,[3-5] feeding activities,[6-8] mating and other social processes.[9,10] Contaminant-induced behaviours have demonstrated rapid response onset and sensitivity at environmentally relevant micropollutant concentrations.[11-14] Aquatic organism movement behaviour from onset micropollutant contamination can take the form of active avoidance like swimming[15-16] or drifting.[16-18] Some organisms such as gastropods can passively avoid the chemical threat by withdrawing in the shell,[19] drifting,[16] or clamping to the substrate,[20] characterized by a reduction or a complete cessation of movement. Locomotor hyperactivity can be linked to a potential ability to exhibit fleeing response – defined here as an instantaneous increased locomotor activity within a few minutes time-frame after exposure, to differentiate from the longer-term delayed-induced avoidance process.[e.g., 15] However, while fleeing behaviour may reduce chemical exposure in the short term, it was associated with costs related to physiological and energetic expenditure,[21] lost feeding opportunities,[22] deleterious consequences of displacement in suboptimal habitats,[23] or risk-taking,[24] with consequences in population dynamics.[25,26] The videotracking of motor behaviours, and particularly the increase in locomotor activity, proved to be a reliable, sensitive method to detect surges of micropollutants such as trace metals, pesticides and pharmaceuticals for both laboratory biotests and real-time biomonitoring of effluents.[14,27] The intensity of the activity response correlated with the concentration of the tested contaminant,[14] and the type of response (hyperactivity, hypoactivity or no response) differed between contaminants[27] and species.[17,28] From an ecological risk assessment perspective, this suggests that contaminant-induced behavioural perturbations, in particular locomotor hyperactivity and fleeing responses, may constitute an effect pathway contributing to chemical hazard additionally to toxicity, which shall vary considerably with compounds and species.

Ecological risk assessment is currently based on outcomes from conventional toxicological tests focusing on mortality (e.g., median lethal concentration, LC50) and chronic sub-lethal effects on development or reproduction (e.g., median effect concentration, EC50). The need for complementary tests to determine threshold values effectively protecting all ecological functions in the environment, including spatial occupancy, is mandatory,[11,26] especially if there is no correlation between their respective outcomes. As an example, some compounds such as herbicides have high LC50 values in animal species including crustaceans, well above environmental concentrations, but trigger behavioural responses in these non-target aquatic species at low concentrations.[29] Current ecological risk assessment procedures rely on the use of available toxicological endpoints, mainly LC50s for many compounds, to rank species according to their sensitivity to a given contaminant, and to describe species sensitivity distribution (SSD).[30,31] For instance, acute-SSDs calculated with acute LC50 values are used in pesticide risk assessment to quantify the potentially affected fraction of species (PAF) when a community is exposed to a given level of environmental concentration,[30] or to derive environmental protection thresholds (e.g., regulatory acceptable concentrations RACs) in order to preserve communities from acute effects of contaminants.[32] But, regarding the possibility of additional ecological impacts mediated by behaviour perturbation, PAF calculation could be flawed if there are discrepancies between species rankings along SSDs either based on mortality endpoints or based on behavioural responses.

Using a large panel of aquatic invertebrate taxa, this study aimed to test the hypothesis of a positive correlation between concentrations thresholds that induce species behavioural locomotor response and traditional toxicity endpoints (LC50s). Our hypothesis was that tolerant species should not activate movement response, while species that are particularly vulnerable to the toxic action of a contaminant may have developed locomotor reactions, either hyperactivity or hypoactivity. Considering 31 invertebrate species (19 freshwater species, 12 marine species), we tested this hypothesis for two substances of interest for aquatic ecosystems, known to induce avoidance in invertebrates: the insecticide imidacloprid and the metal copper. Imidacloprid, and neonicotinoids in general, are widespread systemic insecticides acting as neurotoxicants.[33] Imidacloprid is well-known to adversely affect the survival of non-target aquatic invertebrates, and induces an increase of drift of aquatic invertebrates in flowing waters.[17,18] The SSD approach has clearly established that among freshwater invertebrates, insects are the most sensitive taxa in terms of mortality, followed by crustaceans and finally molluscs.[34] Copper on the other hand, an essential element for all life forms, is a constitutive element of hemocyanin, the respiratory protein in several arthropods and molluscs.[35] Economically important, this historical contaminant of industrial and urban environments is also widely used as antifungal and antiparasitic agent, especially in organic agriculture[36] and aquaculture.[37] Copper has well documented toxic effects on aquatic organisms – with strong variability in taxa sensitivities.[38] It causes spatial avoidance in freshwater shrimps,[39] mayflies,[40] chironomids,[41] snails[42] and leeches,[43] and induces a strong, fast fleeing response in *Gammarus fossarum*.[14] Both substances are growing ecotoxicological concerns for aquatic ecosystems, related to their worldwide increasing use, with peak exposures by agricultural runoff and urban effluents. We firstly assessed the presence of a locomotor response of 16 non-target freshwater species to imidacloprid using the videotracking device ToxMate.[14] Based on our hypothesis and on the well-documented SSDs for freshwater invertebrates towards neonicotinoid lethality,[34,44] we expected that at least some insect species would flee the imidacloprid immediately after exposure, contrary to gastropods or annelids, which are known to survive to high levels of insecticide exposure. Then, we used copper as contaminant and measured both the locomotor responses (behavioural sensitivity) and lethality thresholds (toxicological sensitivity) by acute dose-response tests for a larger range of freshwater or marine invertebrate species (19 freshwater species, 12 marine species). Comparison of the behavioural and toxicological sensitivity rankings of the species for the two substances studied led us to discuss how far mortality data may be insufficient for predicting the relative vulnerability of species community to micropollutant hazard.

# 3. Materials and methods

## 3.1 Species collection

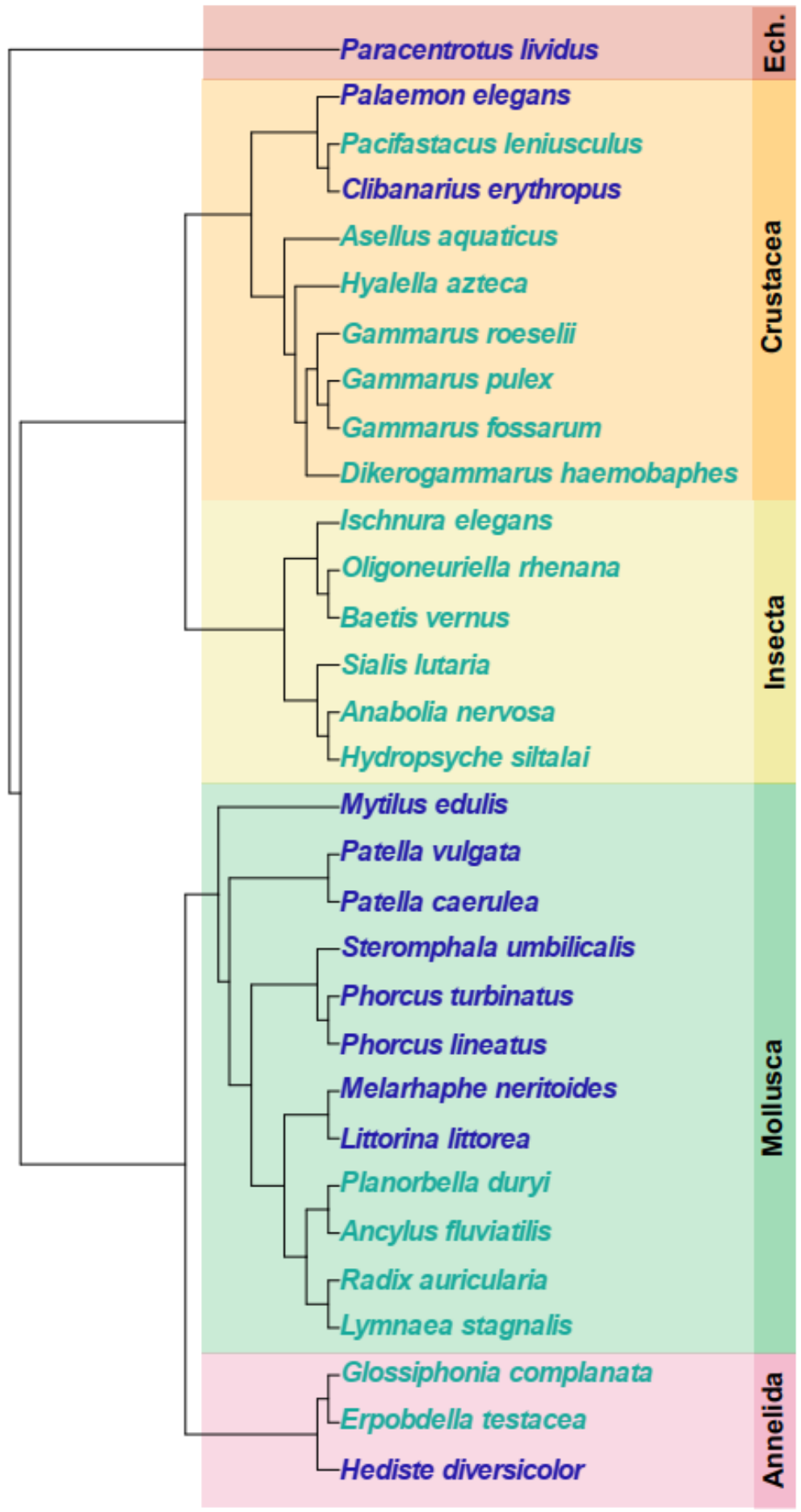

**Figure 1:** Phylogenetic tree of the 31 invertebrate species used in the study. Species in dark blue are marine species, Species in light blue are freshwater species. “Ech.” stands for “Echinoderma”. Location of source populations are detailed in Supplementary Tab. 1.

All experiments combined, a total of 31 species of aquatic invertebrates — 19 freshwater species, 12 marine species — were selected, mainly based on the availability of field populations or in in-house laboratory breeding cultures (Fig. 1 and Supplementary Tab. 1). The individuals were collected from their respective habitat with kick nets (e.g. *Baetis)*, kick nets and sieves (e.g. *Asellus, Erpobdella*, *Gammarus*), or handpicked (e.g. all gastropoda, *Clibanarius)*. Larval stages were used for freshwater insects, and juvenile specimens were selected for species with larger adult body size (urchins, mussels, crayfish, and shrimp), taking into account the dimensions of the observation chambers in the ToxMate videotracking device. Freshwater species were brought to the lab and maintained in well water (with constant aeration, water temperature ~ 12°C, conductivity σ ~ 0.5 mS $cm^{-1}$, water hardness ~ 230 mg $L^{-1}$ $CaCO_3$), and habituated at the lab conditions at least two days before experiments. Marine organisms were housed in filtered seawater obtained from the vicinity of the sampling sites and acclimatised for at least two days prior to experiments.

## 3.2 Behavioural assays

### 3.2.1 Protocol

The behavioural response to copper or imidacloprid (locomotor reaction to contaminant spike) was assessed with the video-tracking biomonitoring station ToxMate (ViewPoint, Civrieux, France), following the design described in Ruck et al.[14] tracking simultaneously up to 48 organisms isolated in individual arena. The observation chambers are hydraulically connected to a central water reservoir via pumps, forming a closed 15 L recirculating system used for the experiments. Each individual arena measures 55 x 50x 18 mm and receives water at a flow rate of 0.35 L per minute. A minimum of 8 individuals (with body size compatible with the arena volume) were used for each concentration of each substance at the beginning of the experiment. The aerated water circulating in the arenas was either natural groundwater drilled within the INRAE lab facilities for freshwater species (pH= 7.7, conductivity σ = 0.49 mS $cm^{-1}$, salinity 2) or seawater from sampling sites (Mediterranean Sea: pH= 7.8, conductivity σ = 56.7 mS $cm^{-1}$, salinity 37.7; The English Channel sea: pH= 8, conductivity σ = 53.1 mS $cm^{-1}$, salinity 32). After two days of laboratory acclimatisation, individuals were placed in the ToxMate, 24h prior the video recording for behavioural analysis. The introduction of the contaminant (i.e., copper or imidacloprid), hereinafter "spiking", took place 12h after the start of the recording. A stock solution of each contaminant was prepared from copper (II) sulfate 5-hydrate ($CuSO_4 \cdot 5H_2O$) or Imidacloprid ($C_9H_{10}ClN_5O_2$; Sigma-Aldrich, Darmstadt, Germany). A determined quantity of the stock solution prepared in deionized water was drawn to reach one of the desired test concentrations — 0.5, 2, 8, 32 and 128 µg $L^{-1}$ for copper, $10^{-2}$, $10^{-1}$, 1, 10 and 100 µg $L^{-1}$ for Imidacloprid. The exposure system was strictly the same as the design presented in the study of Ruck et al. (2023).[14] To prevent an artificially high initial concentration peak - which could have a disproportionate impact on organism health - the appropriate volume of stock solution was first pre-diluted in a beaker containing 300 mL of water drawn from the system. This step ensured rapid and homogeneous mixing throughout the entire volume in the reservoir tank.

### 3.2.2 Behavioural data analysis

*Raw data processing* - The videotracking was realised with the inbuilt ToxMate software (VpCore2 V5.15, ViewPoint Life Sciences, Lyon, France), where raw data (i.e. cartesian coordinates of the centre of gravity of the tracked organism, recorded at 10 fps) is automatically transformed into aggregated distance travelled by each individual over 20 seconds (i.e. 3 data points per minute). Before any statistical treatment, the individual trackings were processed for noise and outliers due to tracking failure producing aberrant measures of travelled distances. The presumed dead individuals because immobile 30 min after the spiking time or earlier, characterized by the absence of movement during 1000 consecutive data points, were automatically detected and removed from behavioural analysis. The test condition was kept in subsequent behavioural analyses if the response at the considered concentration was described by at least four live organisms. The tracking dataset (individual records) is provided as Supplementary Material ("TrackingData.txt").

*Individual response score* - A probabilistic score of response to the spiking (activation/inhibition of locomotor activity) was calculated for each individual. First, the median value of the distance travelled by each individual in 20 seconds within the 10 min after the spiking time (i.e. 30 data points) was calculated to quantify post-spiking activity. The 10-minute observation window was chosen to ensure that, for the species retained in the final dataset, locomotor activity could be reliably recorded under control conditions - particularly during the pre-spiking period. Following several data analysis trials, 10 minutes proved to be a suitable compromise: short enough to maintain reaction specificity to the spiking event, yet long enough to provide a robust and reliable measurement of locomotion. Second, a reference distribution of median values of 30 data points expected in uncontaminated conditions (pre-spiking period) was built for each individual by 999 random resampling with replacement of 30 data points from the pool of distance values recorded within 1h before the spiking time. The ranking of the median value post-spiking – as a 1000$^{th}$ value – in this pre-spiking reference distribution was evaluated (considering the average position in case of ties), providing a rank between 1 and 1000, and then transformed to be bounded between 0 and 1, hereinafter designated as "individual response score" (IRS) to the contaminant. Individual outliers were highlighted, for each species and concentration, by calculating the modified z-scores based on the median values (Iglewicz & Hoaglin, 1993):

$$Modified\ z - score = 0.6745\ |(x_i - \tilde{x})|/MAD$$

with $x_i$ the i$^{th}$ IRS value of the set, $\tilde{x}$ the median IRS value of the set, and MAD the median absolute deviation of the set. Data points with an IRS value outside of the range of the median value +/- 0.05 (i.e. 5% of 1, the maximum reachable value), and a modified z-score greater than 3 were considered as outliers and removed from the analysis.[45] When MAD=0 (i.e. more than half of the points have the same IRS value), every IRS even slightly different from the median is considered as an outlier; the first condition of range membership solved this issue by keeping values close to the median, yet not equal. To ease the interpretation, the IRS after spiking, bounded between 0 and 1, was centred and reduced to be bounded between -1 and 1, 0 meant no change in locomotor activity after spiking compared to the distribution before spiking (Fig. 2a), 1 meant a strong activation of activity (Fig. 2b), and -1, a strong inhibition of activity (Fig. 2c).

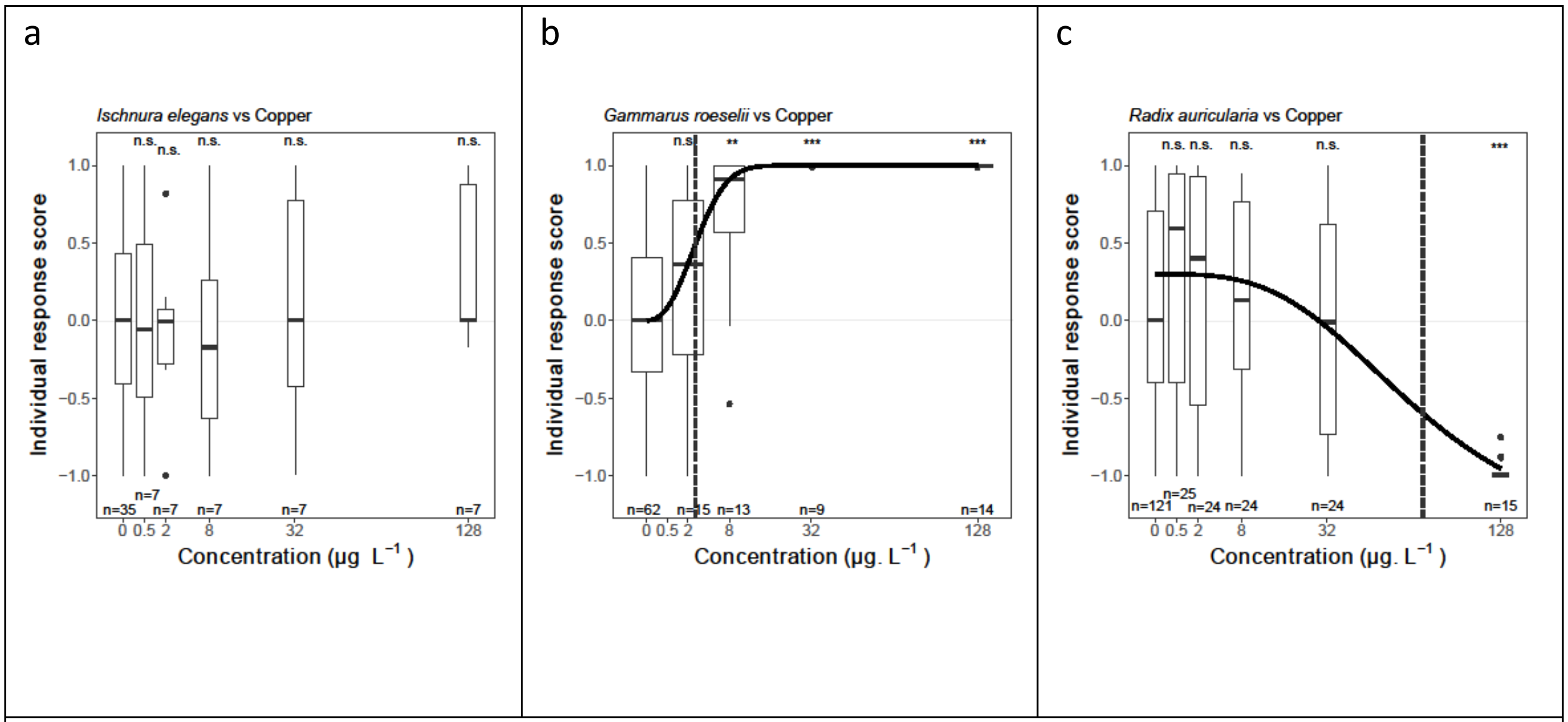

**Figure 2:** Examples of no-response (a), induction (b), and inhibition (c) of locomotor activity as quantified by the change in IRS distribution with increasing spiked copper concentrations. n= number of organisms per concentration; Results of statistical significance test (difference with the pseudo-control distribution) are noticed above each condition; EC50$_{behaviour}$ is symbolized by the vertical blue line for responsive species.

*Dose-response modelling* - A set of pseudo-control IRS values, at concentration 0 µg $L^{-1}$, was created by merging the IRS values calculated from the 10min-period before spiking, for all individuals of each species, irrespective of the tested concentration (with application of the same outliers filtering procedure as post-spiking data). Then, the difference between IRS values in the pseudo-control condition and in each concentration was assessed by a Dunnett test[46] realized with *DescTools* package.[47] To ensure the independence of the data between groups in the comparison between a given concentration and the pseudo-control, the individuals of that concentration were removed from the control data set. Significantly different values from the control, with monotonic or unimodal changes of IRS with increasing concentration of contaminant, were interpreted as a response to the contaminant: species exhibiting this significant response will be designed therefrom "Responsive species", while species with no significant change of activity within the range of tested concentrations of contaminant will be designed as "Unresponsive species". For responsive species, the median IRS values of each concentration were fitted with a dose-response function, determining the best fit between 2- parameter log-logistic function, 4-parameter log-logistic function and Weibull functions by their Akaike Information Criterion (AIC), realized with *drc* package.[48] In case of apparent unimodal response, the first part of the response (and so, monotonic), either increasing or decreasing, was considered in the fitting. An effective concentration $EC50_{behaviour}$ was determined for each combination of responding species by the inflexion point of the functions. The confidence intervals of the $EC50_{behaviour}$ were determined from a set of 1000 dose-response inflexion points applied on same-sized random samples with replacement of IRS in each concentration. The lower and upper limits of the confidence interval were determined as the $25^{th}$ and $975^{th}$ sorted inflexion point values.

The R packages *dplyr*,[49] *drc*,[48] *data.table*,[50] and *googledrive*[51] were used for data management and analysis, and *ggplot2*[52] for visualization.

## 3.3 Toxicity tests

### 3.3.1 Protocol

Because differences in lethal sensitivity to neonicotinoid insecticides are well-described between major invertebrate clades in the literature,[34,44] we focused the toxicity tests to characterize between species variability in threshold lethal concentrations for copper only. Hence, the toxicological sensitivity was assessed through the mortality of the individuals in 96h static toxicity tests along a gradient of copper at 0 (control group), 8, 32, 128, 512 and 2048 µg $L^{-1}$ (four sessions for freshwater species and four sessions for marine species). These nominal concentrations were checked by ICPMS analyses at four different dates of freshwater experiments and one date for marine tests (Supplementary Fig. 1). The limits of quantification were 0.2 and 5 µg $L^{-1}$ for copper in freshwater and marine waters respectively. The sensitive species *Baetis* was also tested at an additional concentration at 4 µg $L^{-1}$. The resistant species with high survivals at 2048 µg $L^{-1}$ were tested in complementary experiments at higher concentrations: 3, 4, 6, 8, 12, 16, 33 and 98 mg $L^{-1}$. These high, environmentally-irrelevant concentrations aimed only to rank the species sensitivities. The tests took place in 500mL plastic beakers, in 12°C- temperature-controlled water bath, in triplicates per concentration. The solutions were obtained by successive dilutions of a stock solution made from copper (II) sulfate 5-hydrate in the same water medium than behavioural tests. Individuals used in these toxicity assays were collected at the same time in the same places as the individuals used in behavioural tests, and have experienced the same environment. A determined number of randomly-chosen individuals (see Supplementary Tab. 2 for details by species) was introduced in the beakers without feeding nor aeration. The number of individuals varied depending on organism availability for each species, the decision to maintain a fixed number of exposure replicates, and the requirement of a minimum of 12 individuals per concentration condition (an exception was made for the echinoderm). Pilot experiment and experimental controls demonstrated the survival of the species under the test conditions at the density used for each species (Survival in control = 0.98 +/- 0.02 across all taxa; Supplementary Tab. 2). The exposure media were renewed after 48h, the survivors counted and the dead individuals removed. After 96h, the remaining individuals were removed, the survivors counted then sacrificed.

### 3.3.2 Mortality data analysis

A reduced general unified threshold model of survival GUTS[53] modelled the mortality in toxicity tests, providing a LC50 value and associated 95% confidence interval for each species at 96h, using *morse* V.3.3.2 package.[54]

## 3.4 $EC50_{behaviour}$ *vs.* LC50 Comparisons

The overall change of ranking of the species based on their respective untransformed LC50 and $EC50_{behaviour}$ (considering only the responsive species) was tested with a Spearman's rank correlation test, for freshwater species and marine species separately.

# 4. Results

## 4.1 Behavioural tests

### 4.1.1 Test of imidacloprid on freshwater species

Out of the 16 tested species, 10 species were unresponsive to imidacloprid: no significant differences were found between the locomotor activity in controls (i.e. 10min-period before spiking) and the 10min-period after spiking at any peak concentrations (up to 100 µg $L^{-1}$), regardless the taxonomic group (Fig. 3 and Supplementary Fig. 2). As peculiar case, *Radix* exhibited a decreasing activity at one intermediate concentration but not at the highest concentration of imidacloprid; unsure of the link between the response and the contaminant, we considered *Radix* as unresponsive in further analyses. Besides, the responsive species were the crayfish *Pacifastacus* ($EC50_{behaviour}$ = 1.22 µg $L^{-1}$), then the damselfly *Ischnura* ($EC50_{behaviour}$ = 10.3 µg $L^{-1}$), the leech *Glossiphonia* ($EC50_{behaviour}$ = 22.2 µg $L^{-1}$), the caddisfly *Hydropsyche* ($EC50_{behaviour}$ = 35.4 µg $L^{-1}$), the demon shrimp *Dikerogammarus* ($EC50_{behaviour}$ = 59.1 µg $L^{-1}$), and the limpet *Ancylus* ($EC50_{behaviour}$ = 64.1 µg $L^{-1}$).

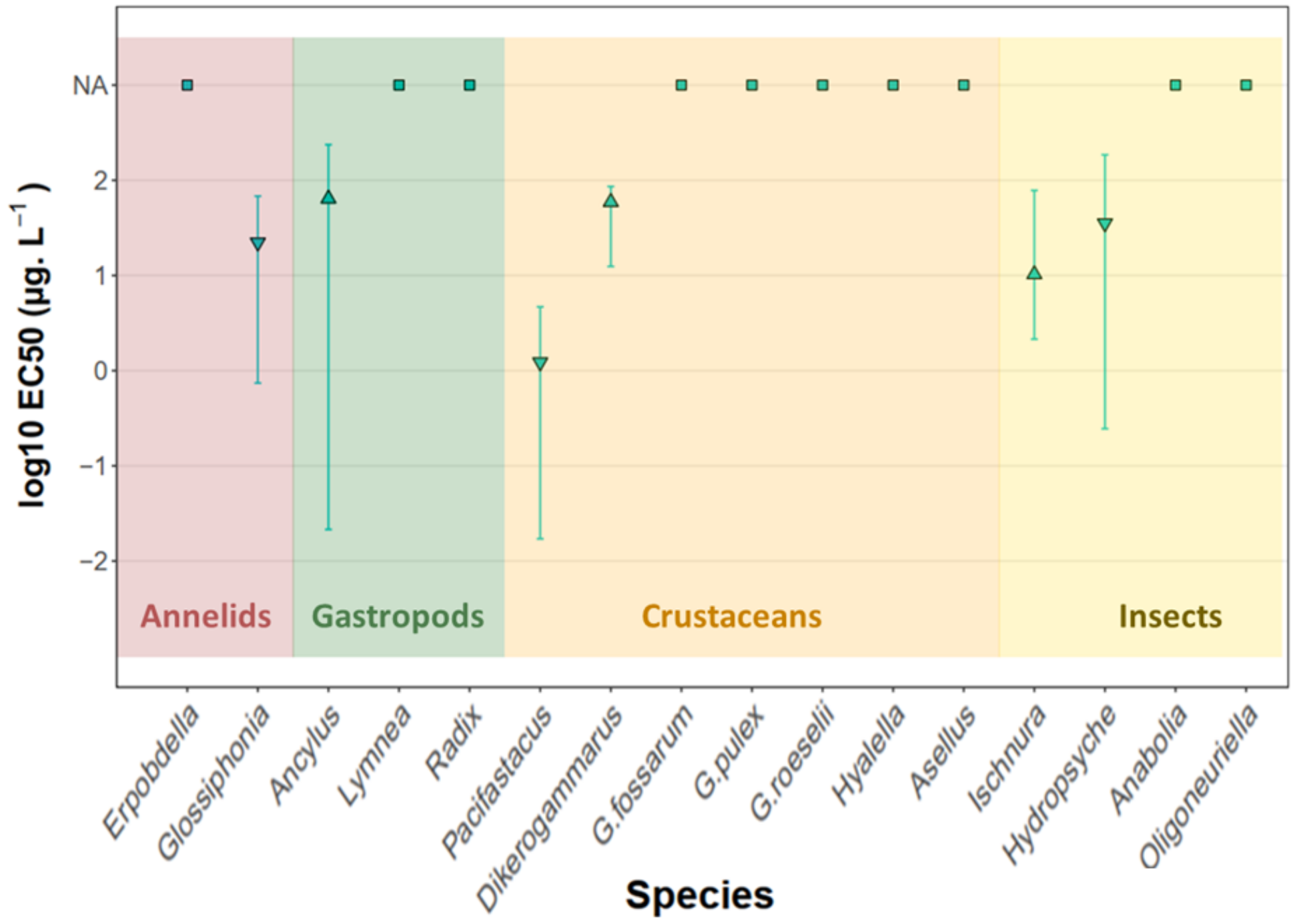

**Figure 3:** $EC50_{behaviour}$ and associated 95% confidence interval for freshwater invertebrates facing imidacloprid spikes. Species include Hirudinae annelids (pink area), Gastropoda (green area), Crustaceans (orange area) and Insecta (yellow area). Upward triangles indicate increasing activity (i.e. hyperactivity), downward triangles indicate decreasing activity (i.e. inhibition response), squares indicate unresponsive species.

### 4.1.2 Test of copper on freshwater and marine species

Out of the 31 tested species, 12 species (6 freshwater species, 6 marine species) were unresponsive to copper spikes (up to 128 µg $L^{-1}$): no significant differences were found between the activity in controls (i.e. 10min-period before spiking) and the 10min-period after spiking regardless of the concentration (Fig. 4 and Supplementary Fig. 3), like the mayfly *Baetis* and four of the marine Gastropoda species. Only the freshwater snails *Lymnaea* and *Radix* exhibited decreasing responses with increasing copper concentrations (See Supplementary Fig. 3). The most sensitive freshwater species exhibiting a change of activity were the amphipod *Dikerogammarus* ($EC50_{behaviour}$ = 0.15 µg $L^{-1}$), the isopod *Asellus* ($EC50_{behaviour}$ = 0.7 µg $L^{-1}$) and the amphipod *Gammarus pulex* ($EC50_{behaviour}$ = 0.75 µg $L^{-1}$), while some species responded only to higher concentration such as the mayfly *Oligoneuriella* (EC50 = 27 µg $L^{-1}$), the snail *Radix* (72.6 µg $L^{-1}$) and the caddisfly *Anabolia* ($EC50_{behaviour}$ = 108 µg $L^{-1}$). The most sensitive marine species were the rockpool shrimp *Palaemon* ($EC50_{behaviour}$ = 0.52 µg $L^{-1}$), the other responsive species exhibiting close values of $EC50_{behaviour}$, between 8.7 and 23 µg $L^{-1}$.

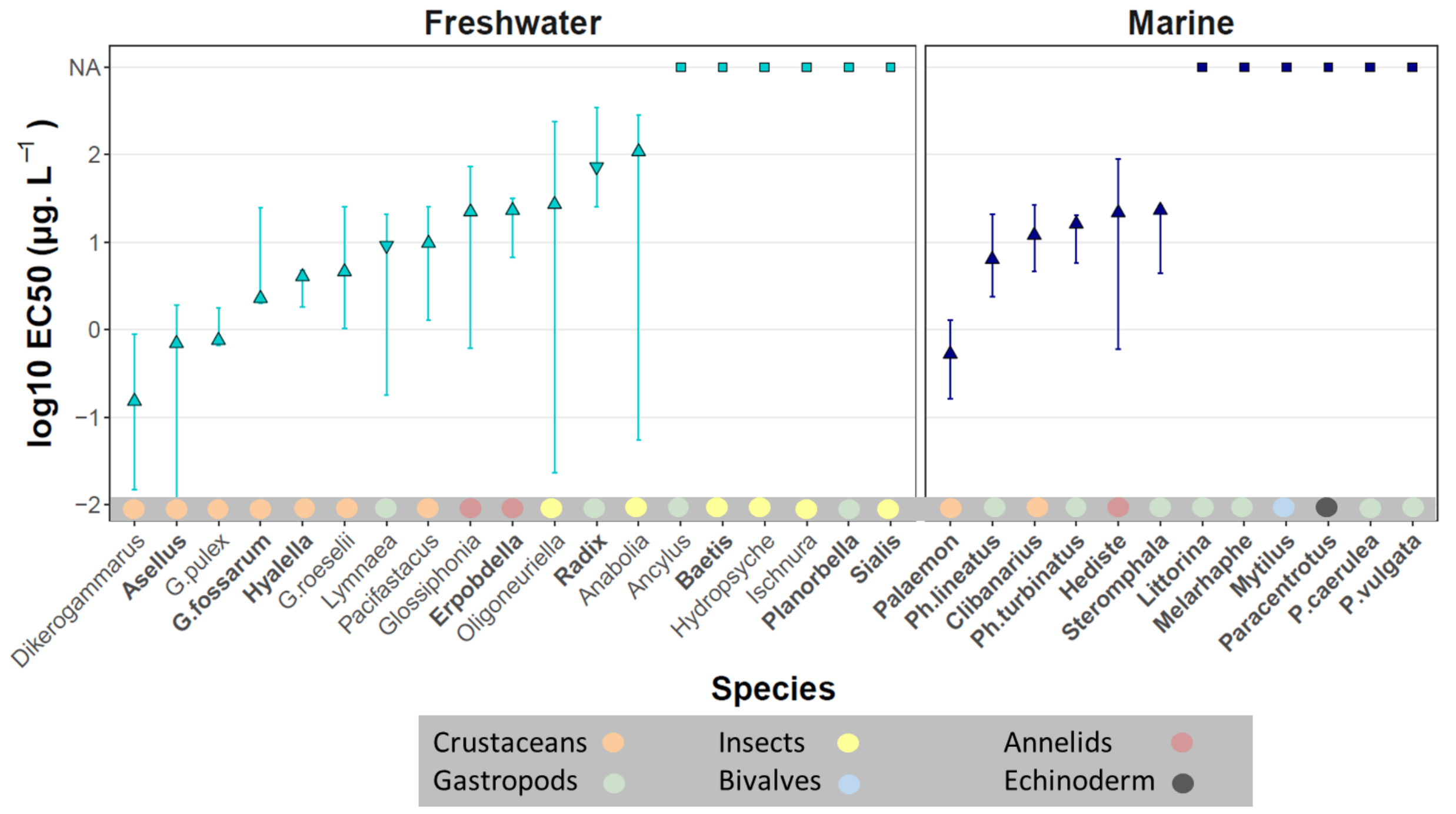

**Figure 4:** $EC50_{behaviour}$ and associated 95% confidence interval for freshwater and marine invertebrates facing copper spikes. Upward triangles indicate increasing activity (i.e. hyperactivity), downward triangles indicate decreasing activity (i.e. inhibition response), and squares indicate unresponsive species. Species tested in copper toxicity assays appear in bold.

## 4.2 Copper toxicity tests

The mortality of every species exceeded 50% within the range of tested copper concentrations, except the freshwater alderfly *Sialis*, who had a 70% mean survival at 98.3 mg $L^{-1}$ of copper after 96h. The most sensitive species was the mayfly *Baetis* (LC50= 17.6 µg $L^{-1}$), followed by the snail *Planorbella* (LC50= 64.7 µg $L^{-1}$) for freshwater species, and the sea urchin *Paracentrotus* (LC50= 30.9 µg $L^{-1}$) for marine species. The most resistant species were *Sialis*, then the freshwater isopod *Asellus* (LC50= 18 mg $L^{-1}$) and the marine rockpool shrimp *Palaemon* (LC50= 5.76 mg $L^{-1}$; Fig. 5).

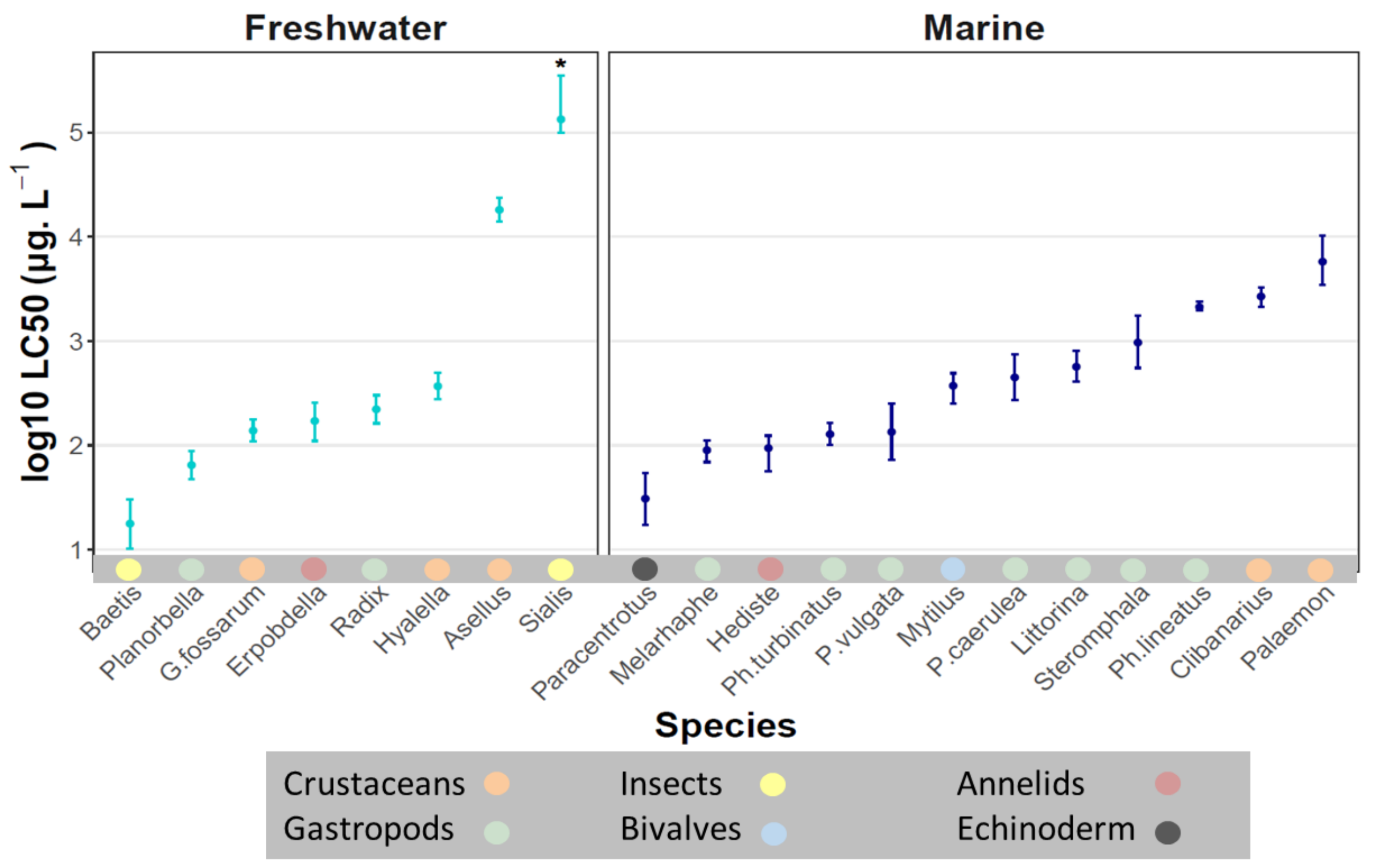

**Figure 5:** Copper 96h-LC50 and associated 95% confidence interval for freshwater and marine invertebrates. The asterisk indicates extrapolated LC50 laying outside of the tested range of copper concentrations.

## 4.3 Relationship between species rankings based on behavioural and toxicological sensitivities to copper

No correlation was observed in the rankings between LC50 and $EC50_{behaviour}$ of responsive freshwater species (Spearman's rank correlation test, $\rho$ = -0.3, *p-value* = 0.68). An illustration is the isopod *Asellus*, species among the most behaviourally sensitive species while it was categorized as resistant species in mortality tests (Fig. 6). The absence of relationship between toxicological ad behavioural responses is also supported by the fact that the three unresponsive freshwater species in behaviour tests are the two highest sensitive species (*Baetis, Planorbella*) in terms of mortality, together with the most tolerant species (*Sialis*). No significant correlation was found either between LC50 and $EC50_{behaviour}$ rankings of responsive marine species (Spearman's rank correlation test, $\rho$ = -0.77, *p-value* = 0.1). Some moderately sensitive species in mortality tests, like the urchin *Paracentrotus*, the snail *Melarhaphe* or the limpet *Patella vulgata*, did not react to copper spikes (Fig. 6). In contrast, the most resistant marine species in the toxicity assays, the shrimp *Palaemon*, was the most sensitive species in the behavioural tests, exhibiting an increasing activity at low concentration of copper spikes. The six unresponsive species in behaviour tests are scattered throughout the median range of species toxicological sensitivities regarding the LC50s.

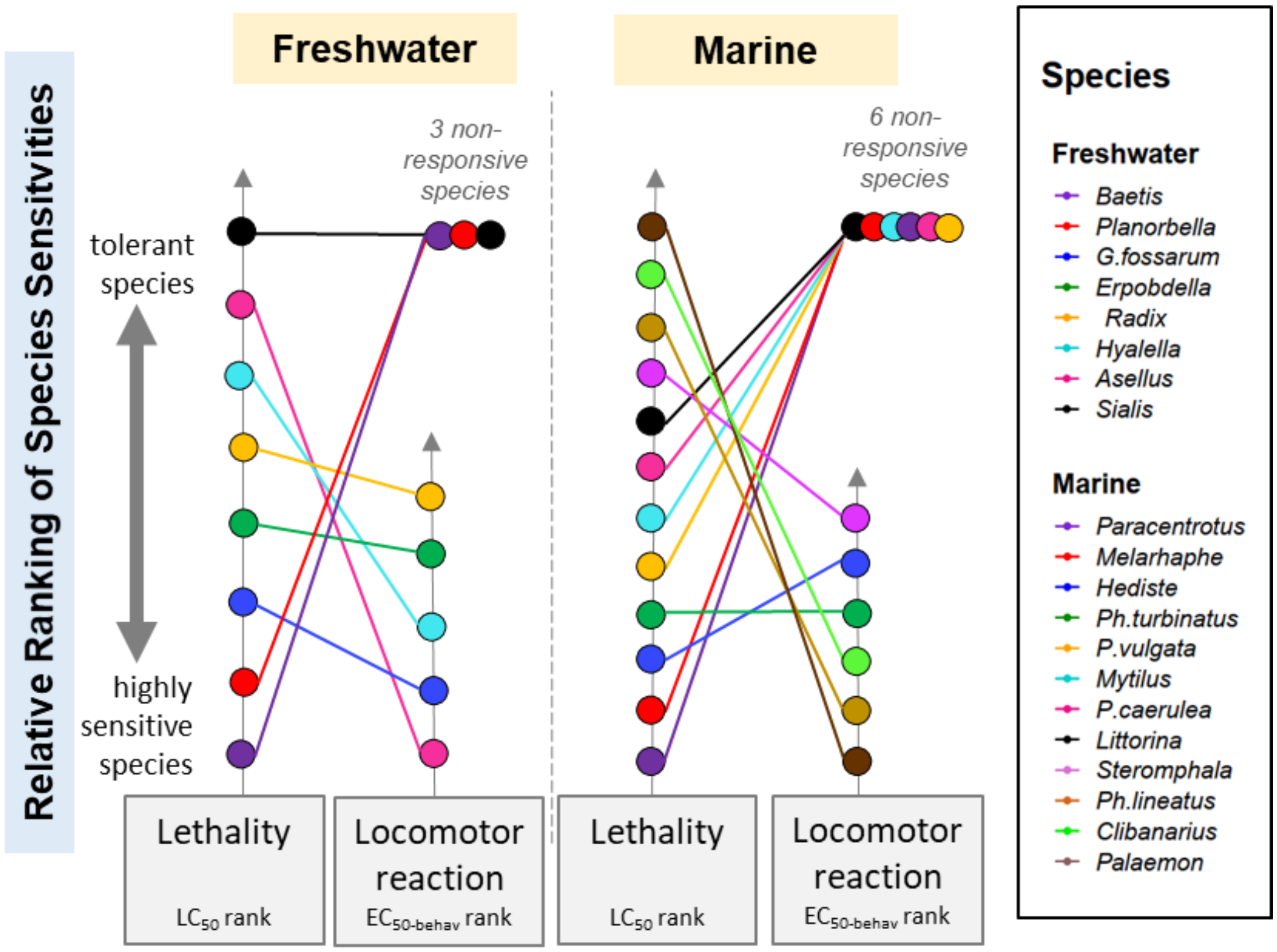

**Figure 6:** Comparison of species sensitivity rankings based on lethality (ranks of 96h-LC50) and locomotor reaction (ranks of $EC50_{behaviour}$ for hyperactivity or locomotor inhibition response in 10min spiking experiments) for copper in freshwater and marine species. Non-responsive species correspond to species with no observed behavioural changes up to 128 µg $L^{-1}$.

# 5. Discussion

In our study, we measured the behavioural responses of several aquatic species to imidacloprid and copper, and their toxicological sensitivities to copper. Tested concentrations of imidacloprid spikes induced a locomotor response to 6 non-target freshwater species from different taxonomic groups, out of 16 tested species. Copper induced such a response in 13 freshwater species out of 19 tested species, and in 6 marine species out of 12. The ranking of behavioural *vs* toxicological sensitivities appeared non-correlated.

## 5.1 Toxicological sensitivities of freshwater and marine species to copper

### 5.1.1 Freshwater species

The 96h LC50s of copper of freshwater invertebrates are consistent with those reported in previous studies on the same species or related taxa (Supplementary Tab. 3). The most sensitive species was the Mayfly *Baetis*, which exhibited sensitivity similar to the other Baetidae *Centroptilum triangulifer*.[55] Some of our LC50 exceeded maximum LC50 found in previous studies (e.g. *Asellus*), possibly related to water conditions during experiments.[56,57] No LC50 was found for *Sialis lutaria* to compare with our value, but its unaffected survival after 8 days at 150 µg $L^{-1}$ confirmed its resistance to copper.[58] The ranking of the toxicological sensitivities from the most sensitive to the most tolerant species was: *Baetis vernus* < *Planorbella duryi* < *Gammarus fossarum* ≈ *Erpobdella testacea* ≈ *Radix auricularia* < *Hyalella azteca* < *Asellus aquaticus* < *Sialis lutaria*. Published relative sensitivities of freshwater invertebrates to copper overall match this ranking, with Baetidae as the most sensitive taxon of our tested species, *Sialis* (when considering its sensitivity to cadmium as metallic contaminant, in absence of data for copper) and *Asellus* as the most resistant, and the rest of the species spread out in the middle ranks.[38] Same rankings were reported with close taxa: *Planorbella trivolvis* < *Gammarus fasciatus* < *Caecidotea (Asellus) intermedia*.[59] A notable difference is *Hyalella*, identified as a sensitive species similar to Baetidae,[38] yet *Hyalella* sensitivity to copper comparable to our finding was previously reported.[57]

### 5.1.2 Marine species

The 96h LC50 for marine species ranged from 31 to 5747 µg $L^{-1}$ of copper, and matched reasonably with reported median lethal concentrations for the same species, but important discrepancies occurred when considering related taxa (Supplementary Tab. 3). *Paracentrotus* was very sensitive to copper, like other sea urchins,[60,61] while crustaceans – the hermit crab *Clibanarius* and the shrimp *Palaemon* – were the most tolerant of our species, as previously reported.[62,63] Reported sensitivities of *Hediste* matched with the upper limit of our confidence interval.[64] Considering its efficient accumulation and detoxification mechanisms,[65] we anticipated a greater resistance to the contaminant from this species. However, the copper sensitivity in *Hediste* is highly variable, with differences of up to one order of magnitude observed between populations.[65] Interestingly, the Gastropoda species exhibited a large range of sensitivities, ranking from second to antepenultimate of our tested species. This finding invites a more detailed examination of the physiological, morphological, and ecological factors that may influence exposure, toxicokinetics, and toxicodynamics underlying the varying vulnerability to copper observed in this marine group. The ranking of the toxicological sensitivities from the most sensitive to the most tolerant species was: *Paracentrotus* < *Melarhaphe* ≈ *Hediste* ≈ *Phorcus turbinatus* ≈ *Patella vulgata* < *Mytilus* ≈ *Patella* *caerulea* ≈ *Littorina littorea* < *Steromphala* < *Phorcus lineatus* < *Clibanarius* < *Palaemon*. As comparison, *Patella vulgata* was also found more sensitive to copper than *Mytilus edulis*.[66]

## 5.2 Locomotor responses to copper and imidacloprid

From a general perspective, the noise observed in locomotor response patterns (notably due to inter-individual variability) could be managed using the developed IRS approach, which allowed us to characterize locomotion sensitivity along concentration gradients. Nevertheless, for certain species and specific trials, the patterns still appear noisy or inconsistent, showing no clear dose-related trend. To advance towards more systematic multi-species comparison studies, it would be advisable to introduce a step for defining validity criteria to determine whether a given species or trial can be included. Such criteria could, for example, be based on a minimum level of locomotion detected under control conditions using a given videotracking device, a minimum variability of locomotion over time, or a more thorough evaluation of the influence of pre- and post-spiking mortality on behavioural indicators (e.g. $EC50_{behaviour}$).

### 5.2.1 Responsive species

Fleeing is an effective response to avoid a toxic environment. Drifts of macroinvertebrates in river increase in presence of imidacloprid.[17,18] We found that some species increases their activity following imidacloprid spiking in Toxmate experiments, as previously observed in similar protocol by *Gammarus fossarum* facing the carbamate insecticide methomyl.[14] The presence or absence of response, as well as the type of response (excitation or inhibition), cannot be associated clearly to the taxonomy: closely related taxa can exhibit different responses, even within a family (e.g. the responsive *Dikerogammarus* vs the unresponsive *Gammarus*). Such interspecific differences in neonicotinoid effects on gammarid locomotion have recently been reported, especially divergent pattern for *Dikerogammarus*.[67] In that study, it is noteworthy that, in contrast to our results (based on reactions measured during 10 minutes after spiking), *Gammarus fossarum* exhibited hyperactivity during a 2-hour exposure to low, neurotoxic concentrations of thiacloprid. Beyond questioning the repeatability of behavioural responses to contaminants across studies, this highlights the importance of experimental design - tailored to the specific objectives of each study (e.g., focusing on immediate reaction versus assessing neurotoxic effects) - in shaping how the results should be interpreted. While the increasing activity recorded in our study for some species confirmed the exposure to the spikes of imidacloprid and justified the categorization of the species as responsive and unresponsive, the noisiness of the individual responses (see Supplementary Fig. 2) cautions against the interpretation of calculated $EC50_{behaviour}$ for imidacloprid as a precise quantitative reproducible metric.

Copper provided clearer response and more reliable $EC50_{behaviour}$. Out of 19 freshwater species and 12 marine ones, respectively 11 and 6 of them increased their activity shortly after copper spiking. Such locomotor hyperactivity was previously recorded in *Gammarus fossarum*,[14] and avoidance was observed in a large range of taxa.[61] The crustaceans were the most sensitive of the tested species in our behaviour assay, as previously observed[11]: all freshwater crustaceans had median effective concentration between 0.15 and 9.7 µg $L^{-1}$, and 0.5 and 12 µg $L^{-1}$ for the marine shrimp *Palaemon* and hermit crab *Clibanarius* respectively. Other crustaceans exhibited such increasing activity to metallic contamination, like Cladocerans, shrimps and crabs, minutes after exposure to copper.[28] Beyond the hyperactivity response observed at very low concentrations in *Palaemon elegans*, no inhibition of locomotor activity was recorded even at the highest copper concentration (Supplementary Fig. 3). Compared to previous forced exposure experiments on *Palaemon* species, this further illustrates the potential for highly diverse behavioural sensitivities within this shrimp genus, including both hyperactivity and hypoactivity responses to copper.[28] Other responsive freshwater species to copper were both leeches, two snails, one mayfly and one caddisfly. For marine species, three snails and the Polychaeta *Hediste* were responsive, with close median effective concentrations (around 10 µg $L^{-1}$). The two responsive freshwater snails, namely *Lymnaea* and *Radix*, reduced their activity after exposure to copper, from respectively 9 and 21 µg $L^{-1}$, but none of the marine gastropod species. Snails exhibit different possible strategies[16] when encountering a threat: fleeing, passively drifting, or isolating itself from the contaminant by retracting in the shell or clamping to the substrate. Drifting and isolating would match the reduction of movement within the 10min after exposure observed in our study. One could argue of a fast toxicity of copper for these species and associate the movement reduction to the death of the individuals. However, the tracking of activity up to 10h after the exposure ensured their survival and the conservation of their moving capacities, unequivocally associating the activity reduction to passive avoidance strategy. Behavioural isolation is an effective, low energy-cost tactic to overcome short-term adverse environmental conditions[68] and to limit the contact with the contaminant.[69]

### 5.2.2 Unresponsive species

A third of the tested freshwater species – including the sensitive mayfly *Baetis*– and half of the marine ones were unresponsive to copper, i.e. exhibiting no change of activity related to copper spikes up to 128 µg $L^{-1}$ within the 10min after exposure. This percentage raised to two third of the tested species to imidacloprid. This apparent lack of reaction to the presence of a toxic substance must be interpreted with caution, given the relatively high variability observed in the data and the potential limitations in statistical power to detect subtle or species-specific responses. Several, non-mutually exclusive, explanations may account for this absence of detectable behavioural reaction. A first explanation could be related to temporality: the individuals did not have enough time to detect the contaminants and to respond adequately, delaying the fleeing response. For instance, the insecticide methomyl induced a gradual, delayed increasing activity in gammarids, peaking 3h after the introduction of the contaminant, despite immediate exposure.[14] A second explanation could be related to an absence of contaminant detection by the individuals. Avoidance response to contaminant is primarily driven by chemoreceptors.[71] This lack of immediate detection may be attributed to the absence or insufficiency of adequate olfactory chemoreceptors. Given the copper data for freshwater species, which point to a general lack of locomotor sensitivity in insect larvae, one hypothesis discussed in the literature is that the life stage may influence sensitivity, as chemoreception capabilities can vary throughout development.[25] This rationale may also apply to urchins and mussels, which were tested at very early juvenile stages due to size limitations of the ToxMate testing arena. This technical constraint may limit our ability to include properly this small set of seemingly non-responsive species in the assessment of interspecies differences. However, a species-level difference still appears to be a non-exclusive likely explanation of behavioural insensitivity, considering the strong between-species variability in copper sensitivity observed among gastropods (all tested at the adult stage) (Fig. 4), as well as the heterogeneous responses to imidacloprid observed in taxonomically related species tested at the same developmental stage (Fig. 3). In addition, at low sublethal concentrations, contaminants as copper can impair the functioning of sensory organs in aquatic decapods.[70,71] This was not illustrated in our case regarding the high behavioural sensitivity of crustacean species to this metal. The last possible explanation lies in the fact that individuals effectively detect the substance, yet exhibit minimal concern toward its presence even for high spiked concentrations (100 µg $L^{-1}$). It is particularly relevant for resistant species like *Sialis*[38] or species efficiently handling copper.[65] For these species, increasing activity and fleeing a contamination they can tolerate without strong deleterious effects would exceed the cost of staying.[72] A decrease in locomotor sensitivity- to copper has already been observed at the intraspecific level in tolerant organisms from populations inhabiting aquatic environments under high contamination pressure.[73] In our case, we did not perform a chemical characterization of water contamination at the collection sites, which limits our ability to draw firm conclusions regarding whether the observed sensitivity differences could be influenced by site-specific effects rather than species identity. However, we took care to ensure that all habitats from which the studied populations were sampled are located in areas without any evident influence from major contamination sources, whether metallic or agricultural.

## 5.3 No correlation between behavioural and toxicological sensitivities: implication for ecological risk assessment

A decorrelation exists between the behavioural and the toxicological sensitivities across both freshwater and marine species datasets. This pattern was first revealed in the experiment with the insecticide imidacloprid, which was selected as a model contaminant due to its higher lethality in arthropods (insects and crustaceans) compared to other animal taxa[34,44,74]: here, the locomotor activity of some insects and crustaceans were unresponsive to spiking, while one gastropod and one annelid species exhibited locomotor reactions. This decorrelation appears also clearly in copper experiments. Our expectation was a positive correlation between lethal concentrations and locomotor response thresholds. Only few species behaved accordingly to this hypothesis, like the resistant, unresponsive *Sialis*, and the sensitive, responsive *Gammarus fossarum*. But the majority of species exhibited locomotor activity changes at concentration levels uncorrelated with lethal thresholds. The first type of such decorrelation is the lack of locomotor responses in toxicologically sensitive species, like the freshwater mayfly *Baetis*, the freshwater snail *Planorbella*, the marine urchin *Paracentrotus* or the marine snail *Melarhaphe*. We did not find any change of locomotor activity in terms of travelled distance, increasing or decreasing, for these species, who were mobile all the time during the videotracking. The lack of immediate locomotor response raises high concerns for these species in case of acute, punctual release of copper, increasing their ecological vulnerabilities. The decorrelation also stands for behavioural responsive species. The freshwater snail *Radix* increased its activity only at high copper concentrations (Fig. 4), while presenting a median level of sensitivity among the SSD based on 96h LC50 (Fig. 5). But the most striking examples of such decorrelation were species displaying locomotor reactions despite inherent resistance to copper exposure, such as *Asellus*, *Hyalella* in some extent, and above all *Palaemon elegans*, the most tolerant marine species tested and yet the first species increasing its activity to copper spike. As comparison with other shrimp species, *Palaemon varians* and the tropical *Penaeus vannamei* avoid copper during 4h exposure tests for concentration in the range of 5 µg $L^{-1}$, while their respective 96h CL50 are >30 mg $L^{-1}$ and 3.9 mg $L^{-1}$.[75,76] These results suggest that the decorrelation between behavioural and toxicological sensitivities may be common, at least in shrimps. Under the hypothesis of potential detrimental consequences of locomotor reactions for organisms fitness or population dynamics, the decorrelation between behavioural and toxicological sensitivities can thus have significant implications for ecosystem functioning and ecological risk assessment. Indeed, a low concentration of copper, considered as harmless in terms of mortality or reproduction for certain species, can induce locomotor response and avoidance reactions, with potential delayed deleterious impacts at the individual and population levels.[73]

The substantial changes in species sensitivity rankings that arise when considering behavioural locomotor responses have important implications for ecological risk assessment. On one hand, our results for copper support the conclusion of Alcívar et al.[11] that environmental protection thresholds derived from toxicological endpoints encompass the sensitivity range of avoidance responses. For instance, applying the EFSA procedure[32] to define a protective concentration for acute effects — using the 96h-LC50 values from our study (Fig. 5) — yields an acute Regulatory Acceptable Concentration (RAC) of 4 µg $L^{-1}$ for marine species and 2 µg $L^{-1}$ for freshwater species (excluding the outlier *Sialis*). These values, calculated as the geometric mean of LC50s adjusted with an assessment factor of 100 (EFSA-Tier 2A for acute effect assessment), closely align with those reported by Alcívar et al.[11] Moreover, these RACs effectively prevent behavioural disturbances, as they remain below the $EC50_{behaviour}$ for most tested species (11/12 marine and 16/19 freshwater species). On the other hand, while incorporating locomotor reactions has minimal impact on overall community protection thresholds, the species ranking decoupling observed in our study (Fig. 6) substantially alters the risk assessment for a given contamination level, under the hypothesis that locomotor reactions translate into detrimental population impacts To illustrate this, we calculated three types of SSDs for copper in marine species (Supplementary Fig. 4): (i) a conventional acute-SSD for mortality risk, based on 96h-LC50 values divided by 10, following Malaj et al.[77]; (ii) an SSD accounting solely for locomotor reactions, based on $EC50_{behaviour}$ values; and (iii) a combined SSD integrating both mortality and locomotor reactions, using the lower of the two values for each species. Once again, protection levels for the invertebrate community remain within the same order of magnitude, regardless of the SSD type. For instance, hazardous concentrations affecting 5% of species range between 0.5 and 2.5 µg $L^{-1}$. However, differences in species ranking between lethality and behavioural sensitivity lead to an underestimation of the potentially affected fraction of species when only mortality is considered. Resistant species such as *Palaemon elegans* and *Phorcus lineatus* are reclassified among the most vulnerable when their escape behaviour is taken into account. PAF is also underestimated if only behavioural sensitivity is considered regardless of mortality. For example, at a concentration of 30 µg $L^{-1}$, 44% of species face mortality risk, 35% exhibit risk of escape reactions, yet the overall PAF rises in reality to 79% when both effects are considered, due to the lack of correlation between species rankings across these two impact types (Supplementary Fig. 4).

In conclusion, the discrepancies between species sensitivity rankings based on either toxicological or behavioural locomotor endpoints call into critical question the estimation of potentially affected fractions of species when acute risk is assessed from a species ranking based solely on toxicological criteria. It is now necessary to evaluate our findings alongside evidence of the real impacts of species' locomotor reactions to contaminants at the population level. It advocates for multi-endpoint approaches, including behavioural locomotor responses in chemical hazard assessment, in order to protect the species and the entirety of ecosystem functions potentially ignored by standard toxicity tests.

# Supporting information

"TrackingData.txt": Raw tracking dataset corresponding to the individual records during pulse exposure experiments.
"Suppl_data.docx": Additional experimental details (species sampling, replication design, exposure conditions), LC50 data from the literature, and dose-response analyses (species-specific behavioural responses, acute SSDs).

# Declaration of competing interest

The authors declare the following financial interests/personal relationships which may be considered as potential competing interests: INRAE and the Viewpoint Company share in co-ownership patents related to the ToxMate technology (patent #WO2018178585; patent #WO2019016273).

## Acknowledgements

This research received financial support from Région Auvergne-Rhône-Alpes (projet Pack Ambition Recherche 2020 - ToxPrints), ANRT (Cifre PhD Grant of GR), institut Carnot Eau-Environnement. The authors would like to thank Mayélé Burlion-Giorgi (SEBIO, Le Havre) and Bastien Chouquet (CSLN, Le Havre) for their help on sampling and identification of marine species. Hervé Quéau, Nicolas Delorme and Rémi Recoura-Massaquant (INRAE, RiverLy) are acknowledged for their support during freshwater species collection and laboratory experiments.

Supplementary material for:

## Rank ordering of species sensitivities to chemicals: discrepancies between behavioural and lethal responses in freshwater and marine invertebrates

Julien Mocq [a*], George Ruck [a b], Michaël Demortier [a], Flavie Brun [a], Romain Coulaud [c], Olivier Geffard [a] and Arnaud Chaumot [a*]

[a] INRAE, UR RiverLy, Laboratoire d'écotoxicologie, F-69625 Villeurbanne, France
[b] Viewpoint, 67 Rue Copernic 01390 Civrieux, France
[c] Université Le Havre Normandie, Normandie Univ, FR CNRS 3730 SCALE, UMR-I 02 SEBIO, F-76600, Le Havre, France
* Emails: julien.mocq@gmail.com ; arnaud.chaumot@inrae.fr

**Content:**

**Supplementary Table 1:** List of species used in behavioural and toxicity assays; location of collected populations.

**Supplementary Table 2:** Details of the replication design followed for the mortality tests (number of individuals).

**Supplementary Table 3:** Comparison of median copper lethal concentration with 95% confidence interval in our study and in (non exhaustive) literature of same or closely related species.

**Supplementary Figure 1:** Nominal vs measured concentrations of copper in freshwater and marine water, in toxicity assays. The horizontal grey lines indicate the Limit of Quantification (LQ) for both types of waters. The solid diagonal represents for the 1:1 correlation.

**Supplementary Figure 2:** Individual Response Score (IRS) distribution along the gradient of imidacloprid peak concentrations (freshwater species). n= number of organisms per concentration; Results of statistical significance test (difference with the pseudo-control distribution) are noticed above each condition; $EC50_{behaviour}$ is symbolized by the vertical blue line for responsive species only.

**Supplementary Figure 3:** Individual Response Score (IRS) distribution along the gradient of copper peak concentrations (freshwater and marine species). n= number of organisms per concentration; Results of statistical significance test (difference with the pseudo-control distribution) are noticed above each condition; $EC50_{behaviour}$ is symbolized by the vertical blue line for responsive species only.

**Supplementary Figure 4:** Acute Species Sensitivity Distributions (SSD) for copper in marine species established on lethality data (LC50/10), behavioural responses ($EC50_{behaviour}$) and combination of both. Log-logistic models were fitted with the MOSAIC software integrating right-censored data for $EC50_{behaviour}$ (Charles et al., 2018).

**References for Supplementary material**

**Supplementary Table 1:** List of species used in behavioural and toxicity essays; location of collected populations.

| Species | Taxonomy | Sampling Site | GPS coordinates | Environment |
|---|---|---|---|---|
| *Anobolia nervosa* | Insecta, Trichoptera, Limnephilidae | Pond along the Vologne river, Laval sur Vologne, FR | 48°11'30"N, 6°41'44"E | Freshwater |
| *Ancylus fluviatilis* | Gastropoda, Planorbidae | Artilla river, Thurins, FR | 45°40'15"N, 4°36'20"E | Freshwater |
| *Asellus aquaticus* | Malacostraca, Isopoda, Asellidae | Le Pollon brook, Saint-Maurice-de-Rémens, FR | 45°57'24"N, 5°15'42"E | Freshwater |
| *Baetis vernus* | Insecta, Ephemeroptera, Baetidae | Rizan brook, Meyzieu, FR | 45°47'14"N, 5°0'49"E | Freshwater |
| *Clibanarius erythropus* | Malacostraca, Decapoda, Diogenidae | Intertidal zone, Le Pradet, FR | 43°05'49"N, 6°01'11"E | Marine |
| *Dikerogammarus haemobaphes* | Malacostraca, Amphipoda, Gammaridae | Rhône river, Cordon, FR | 45°37'29"N, 5°36'57"E | Freshwater |
| *Erpobdella testacea* | Hirudinida, Erpobdellidae | Watercress bed, Saint-Maurice-de-Rémens, FR | 45°57'24"N, 5°15'42"E | Freshwater |
| *Gammarus fossarum* | Malacostraca, Amphipoda, Gammaridae | Watercress bed, Saint-Maurice-de-Rémens, FR | 45°57'24"N, 5°15'42"E | Freshwater |
| *Gammarus pulex* | Malacostraca, Amphipoda, Gammaridae | Suzon river, Val Suzon, FR | 47°24'13"N, 4°52'59"E | Freshwater |
| *Gammarus roeselii* | Malacostraca, Amphipoda, Gammaridae | Bourbre river, Tignieu-Jameyzieu, FR | 45°42'56"N, 5°9'33"E | Freshwater |
| *Glossiphonia complanata* | Hirudinida, Glossiphoniidae | Watercress bed, Saint-Maurice-de-Rémens, FR | 45°57'24"N, 5°15'42"E | Freshwater |
| *Hediste diversicolor* | Phyllodocida, Nereididae | Harbour mud, Le Havre, FR | 49°29'22"N, 0°05'48"E | Marine |
| *Hyalella azteca* | Malacostraca, Amphipoda, Hyalellidae | Lab husbandry, Villeurbanne, FR | | Freshwater |
| *Hydropsyche siltalai* | Insecta, Trichoptera, Hydropsychidae | Le Gier brook, Saint Chamond, FR | 45°26'24"N, 4°30'37"E | Freshwater |
| *Ischnura elegans* | Insecta, Odonata, Coenagrionidae | Pond along the Vologne river, Laval sur Vologne, FR | 48°11'30"N, 6°41'44"E | Freshwater |
| *Littorina littorea* | Gastropoda, Littorinidae | Intertidal zone, Blainville s/ Mer, FR | 49°03'37"N, -1°36'27"E | Marine |
| *Lymnaea stagnalis* | Gastropoda, Lymnaeidae | Seymard river, Saint Maurice de Remens, FR | 45°58'N, 5°16'12"E | Freshwater |
| *Melarhaphe neritoides* | Gastropoda, Littorinidae | Intertidal zone, Hyères, FR | 43°02'03"N, 6°07'37"E | Marine |
| *Mytilus edulis* | Bivalvia, Mytilida, Mytilidae | Harbour bank, Le Havre, FR | 49°29'22"N, 0°05'48"E | Marine |
| *Oligoneuriella rhenana* | Insecta, Ephemeroptera, Oligoneuriidae | Grosne river, Deux-Grosnes, FR | 46°15'46"N, 4°33'3"E | Freshwater |
| *Pacifastacus leniusculus* | Malacostraca, Decapoda, Astacidae | Onzon river, LA Tour en Jarez | 45°28'52"N, 4°23'17"E | Freshwater |
| *Palaemon elegans* | Malacostraca, Decapoda, Palaemonidae | Intertidal zone, Blainville s/ Mer, FR | 49°03'37"N, -1°36'27"E | Marine |
| *Paracentrotus lividus* | Echinoderma Parechinidae | Lab husbandry, Blainville s/ Mer, FR | | Marine |
| *Patella caerulea* | Gastropoda, Patellidae | Intertidal zone, Le Pradet, FR | 43°05'49"N, 6°01'11"E | Marine |
| *Patella vulgata* | Gastropoda, Patellidae | Intertidal zone, Blainville s/ Mer, FR | 49°03'37"N, -1°36'27"E | Marine |
| *Phorcus lineatus* | Gastropoda, Trochidae | Intertidal zone, Blainville s/ Mer, FR | 49°03'37"N, -1°36'27"E | Marine |
| *Phorcus turbinatus* | Gastropoda, Trochidae | Intertidal zone, Le Pradet, FR | 43°05'49''N, 6°01'11''E | Marine |
| *Planorbella duryi* | Gastropoda, Planorbidae | Lab husbandry, Villeurbanne, FR | | Freshwater |
| *Radix auricularia* | Gastropoda, Lymnaeidae | Watercress bed,<br>Saint-Maurice-de-Rémens, FR | 45°57'24''N, 5°15'42''E | Freshwater |
| *Sialis lutaria* | Insecta, Megaloptera, Sialidae | Le Pollon brook,<br>Saint-Maurice-de-Rémens, FR | 45°57'24''N, 5°15'42''E | Freshwater |
| *Steromphala umbilicalis* | Gastropoda, Trochidae | Intertidal zone, Blainville s/ Mer, FR | 49°03'37''N, -1°36'27''E | Marine |

**Supplementary Table 2:** Details of the replication design followed for the mortality tests (number of individuals).

| | Per beaker | Per concentration (µg /L) | | | | | | | | | | | | | | | Mean survival in control (+/- sd) |
|---|---|---|---|---|---|---|---|---|---|---|---|---|---|---|---|---|---|
| | | 0 | 4 | 8 | 32 | 128 | 512 | 2048 | 3072 | 4096 | 6144 | 8192 | 12288 | 16384 | 32768 | 98304 | |
| *Asellus aquaticus* | 10 | 60 | | 30 | 30 | 30 | 30 | 60 | 30 | 30 | 30 | 30 | 30 | 30 | 30 | 30 | 0.95 ± 0.05 |
| *Baetis vernus* | 15 | 45 | 45 | 45 | 45 | 45 | 45 | 45 | | | | | | | | | 0.87 ± 0.12 |
| *Clibanarius erythropus* | 8-10 | 30 | | 30 | 29 | 30 | 30 | 30 | | 24 | 24 | 24 | | | | | 1 ± 0 |
| *Erpobdella testacea* | 6 | 12 | | 9 | 11 | 18 | 22 | 23 | | | | | | | | | 1 ± 0 |
| *Gammarus fossarum* | 10 | 30 | | 30 | 30 | 30 | 30 | 30 | | | | | | | | | 1 ± 0 |
| *Gammarus sp* | 3-4 | 12 | | 12 | 12 | 12 | 12 | 12 | | 12 | | 12 | 12 | | | | 1 ± 0 |
| *Hediste diversicolor* | 4 | 12 | | 12 | 12 | 12 | 12 | 12 | | | | | | | | | 1 ± 0 |
| *Hyalella azteca* | 15 | 45 | | 45 | 45 | 45 | 45 | 45 | | | | | | | | | 0.93 ± 0 |
| *Littorina littorea* | 8 | 24 | | 24 | 24 | 24 | 24 | 24 | | | | | | | | | 1 ± 0 |
| *Melarhaphe neritoides* | 10 | 30 | | 30 | 30 | 30 | 30 | 30 | | | | | | | | | 1 ± 0 |
| *Mytilus edulis* | 8 | 24 | | 24 | 24 | 24 | 24 | 24 | | | | | | | | | 1 ± 0 |
| *Paracentrotus lividus* | 3 | 9 | | 9 | 9 | 9 | 9 | 9 | | | | | | | | | 1 ± 0 |
| *Phorcus lineatus* | 9 | 27 | | 27 | 27 | 27 | 27 | 27 | | 27 | | | | | | | 1 ± 0 |
| *Phorcus turbinatus* | 10 | 30 | | 30 | 30 | 30 | 30 | 30 | | | | | | | | | 1 ± 0 |
| *Planorbella duryi* | 7 | 21 | | 21 | 21 | 21 | 21 | 21 | | | | | | | | | 1 ± 0 |
| *Palaemon elegans* | 6 | 18 | | 18 | 18 | 18 | 18 | 18 | | 18 | | 18 | 18 | | | | 0.94 ± 0.1 |
| *Patella caerulea* | 8 | 24 | | 24 | 24 | 24 | 24 | 24 | | | | | | | | | 0.92 ± 0.07 |
| *Patella vulgata* | 5 | 15 | | 15 | 15 | 15 | 15 | 15 | | | | | | | | | 1 ± 0 |
| *Radix auricularia* | 10 | 30 | | 30 | 30 | 30 | 30 | 30 | | | | | | | | | 1 ± 0 |
| *Sialis lutaria* | 10 | 90 | | 30 | 30 | 30 | 30 | 30 | | 30 | 30 | 30 | 30 | 30 | 30 | 30 | 0.95 ± 0.05 |
| *Steromphala umbilicalis* | 5-6 | 18 | | 18 | 18 | 18 | 17 | 18 | | 18 | | | | | | | 1 ± 0 |

**Supplementary Table 3:** Comparison of median copper lethal concentration with 95% confidence interval in our study and in (non exhaustive) literature of same or closely related species.

| Environment | $Species_{study}$ | 96h $LC50_{study}$ (µg $L^{-1}$) | $LC50_{Literature}$ (µg $L^{-1}$) | Exposure time $_{Literature}$ | Tested species $_{Literature}$ | Reference |
|---|---|---|---|---|---|---|
| Freshwater | *Baetis vernus* | 17.5 [10.2 - 30] | 10.7 | 48h | *Centroptilum triangulifer* (Baetidae) | (Struewing et al., 2015) |
| | *Planorbella duryi* | 64.4 [47.2 - 88.1] | 191<br>320 | 48h<br>96h | *Biomphalaria glabrata*<br>Planorbella trivolvis | (de Oliveira-Filho et al., 2004)<br>(Ewell et al., 1986) |
| | *Gammarus fossarum* | 138.5 [110.1 - 176.5] | [141-2338]<br>[25-1.5.$10^3$] | 96h<br>96h | *Gammarus fossarum*<br>*Gammarus spp.* | (Schmidlin et al., 2015)<br>(U.S. Environmental Protection Agency, 2024) |
| | *Erpobdella testacea* | 171 [109.4 - 257.7] | 200 | 96h | *Erpobdella octoculata* | (U.S. Environmental Protection Agency, 2024) |
| | *Radix auricularia* | 221.2 [160.8 - 302.6] | 172<br>31<br>27 | 96h<br>96h<br>96h | *Radix (Lymnaea) luteola*<br>*Lymnaea stagnalis*<br>*Lymnaea luteola* | (Mathur et al., 1981)<br>(Brix et al., 2011)<br>(Khangarot & Ray, 1988) |
| | *Hyalella azteca* | 368.2 [273.4 - 498] | [4-369]<br>[17-210]<br>344 | 96h<br>96h<br>96h | *Hyalella azteca*<br>*Hyalella azteca*<br>*Hyalella azteca* | (U.S. Environmental Protection Agency, 2024)<br>(Borgmann et al., 2005)<br>(Richards et al., 2015) |
| | *Asellus aquaticus* | 18.1.$10^3$ [14.$10^3$ – 23.8.$10^3$] | 32.$10^3$<br>32.$10^3$<br>9210 | 96h<br>48h<br>96h | *Caecidotea (Asellus) intermedia*<br>*Asellus aquaticus*<br>*Asellus aquaticus* | (Ewell et al., 1986)<br>(U.S. Environmental Protection Agency, 2024)<br>(Martin & Holdich, 1986) |
| | *Sialis lutaria* | 133.4.$10^3$ [99.8.$10^3$ – 336.3.$10^3$] | NA | NA | NA | NA |
| Marine | *Paracentrotus lividus* | 31 [17.2 - 54.9] | 25<br>[22-49] | 96h<br>96h | *Diadema antillarum*<br>*Paracentrotus lividus* | (Bielmyer et al., 2005)<br>(U.S. Environmental Protection Agency, 2024) |
| | *Melarhaphe neritoides* | 89.9 [69.7 - 112.7] | >686- >1487<br>367-398 | 96h<br>96h | *Laevilittorina caliginosa*<br>*Laevilitorina hamiltoni* | (Holan, King, & Davis, 2018)<br>(Holan et al., 2017) |
| | *Hediste diversicolor* | 94.6 [56 - 123.5] | 125<br>[125- 4.1.$10^3$] | 96h<br>96h | *Hediste diversicolor*<br>*Hediste diversicolor* | (Moreira et al., 2005)<br>(U.S. Environmental Protection Agency, 2024) |
| | *Phorcus turbinatus* | 127.8 [101.1 - 161.9] | 286 | 96h | *Cantharidus capillaceus* | (Holan et al., 2017) |
| | *Patella vulgata* | 135.1 [72.4 - 248.5] | 82<br>69 | 96h<br>96h | *Helcion concolor*<br>*Cellana capensis* | (De Pirro & Marshall, 2005)<br>(De Pirro & Marshall, 2005) |
| | *Mytilus edulis* | 373.7 [250.4 - 491.6] | 280<br>480 | 96h<br>96h | *Mytilus edulis*<br>*Mytilus edulis* | (Abel, 1976)<br>(Amiard-Triquet, Berthet et al., 1986) |
| | *Patella caerulea* | 447.1 [271.8 - 748.9] | 82<br>69 | 96h<br>96h | *Helcion concolor*<br>*Cellana capensis* | (De Pirro & Marshall, 2005)<br>(De Pirro & Marshall, 2005) |
| | *Littorina littorea* | 566.3 [403.2 - 800.3] | >686- >1487<br>367-398 | 96h<br>96h | *Laevilittorina caliginosa*<br>*Laevilitorina hamiltoni* | (Holan et al., 2018)<br>(Holan et al., 2017) |
| | *Steromphala umbilicalis* | 976.2 [555.4 - 1756.3] | 286 | 96h | *Cantharidus capillaceus* | (Holan et al., 2017) |
| | *Phorcus lineatus* | 2118.5 [1980 - 2386.6] | 286 | 96h | *Cantharidus capillaceus* | (Holan et al., 2017) |
| | *Clibanarius erythropus* | 2671.7 [2125.2 - 3265.1] | 6500<br>1906<br>8000 | 96h<br>96h<br>96h | *Diogenes pugilator*<br>*Clibanarius africanus*<br>*Clibanarius longitarsus* | (Bat et al., 1999)<br>(Otitoloju et al., 2006)<br>(Lyla & Khan, 2010) |
| | *Palaemon elegans* | 5746.9 [3446 - 10356.1] | 2520<br>[610-3270] | 96h<br>96h | *Palaemon elegans*<br>*Palaemon elegans* | (Bat et al., 1999)<br>(Lorenzon et al., 2000) |

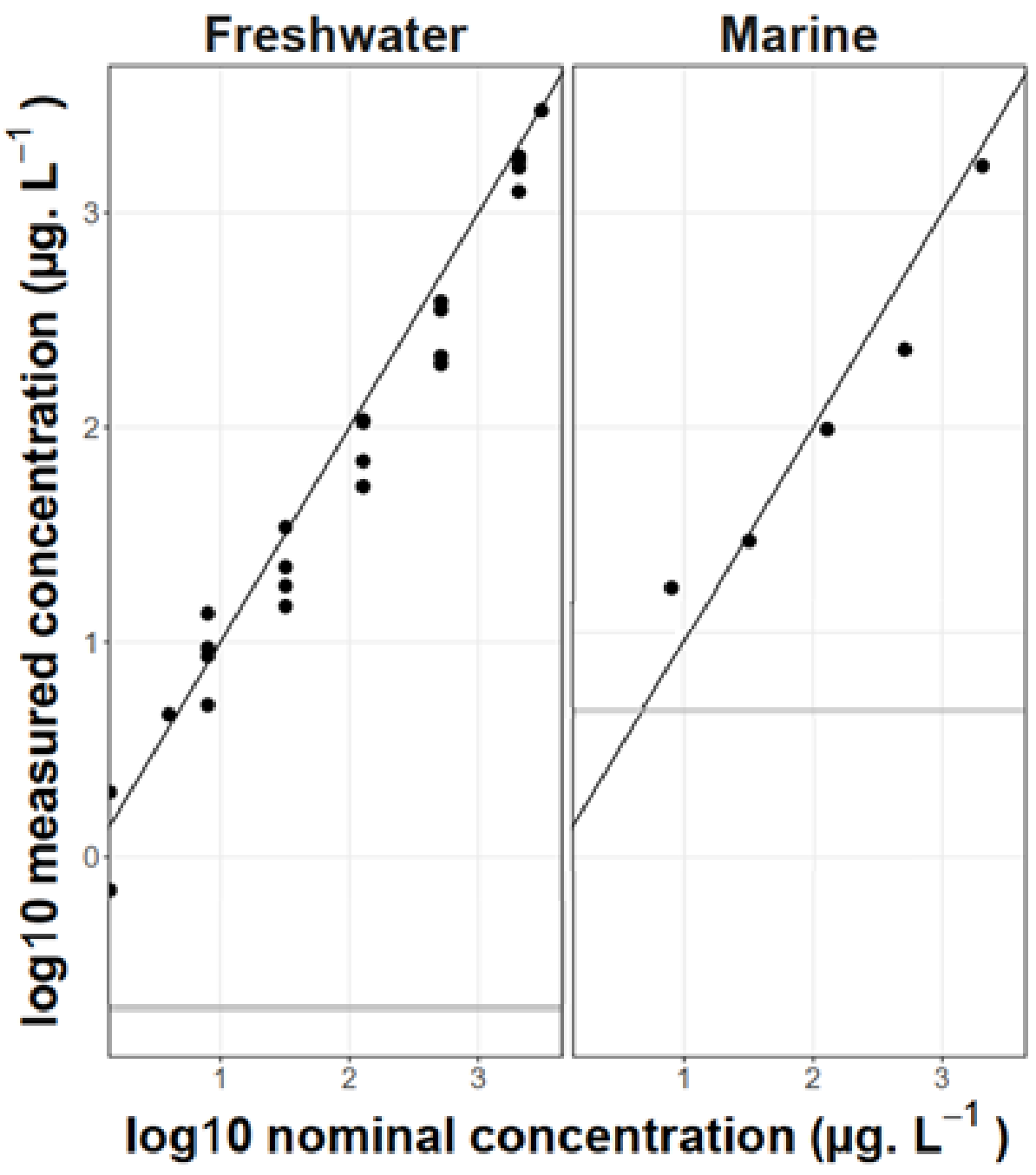

**Supplementary Figure 1:** Nominal vs measured concentrations of copper in freshwater and marine water, in toxicity assays. The horizontal grey lines indicate the Limit of Quantification (LQ) for both types of waters. The solid diagonal represents for the 1:1 correlation.

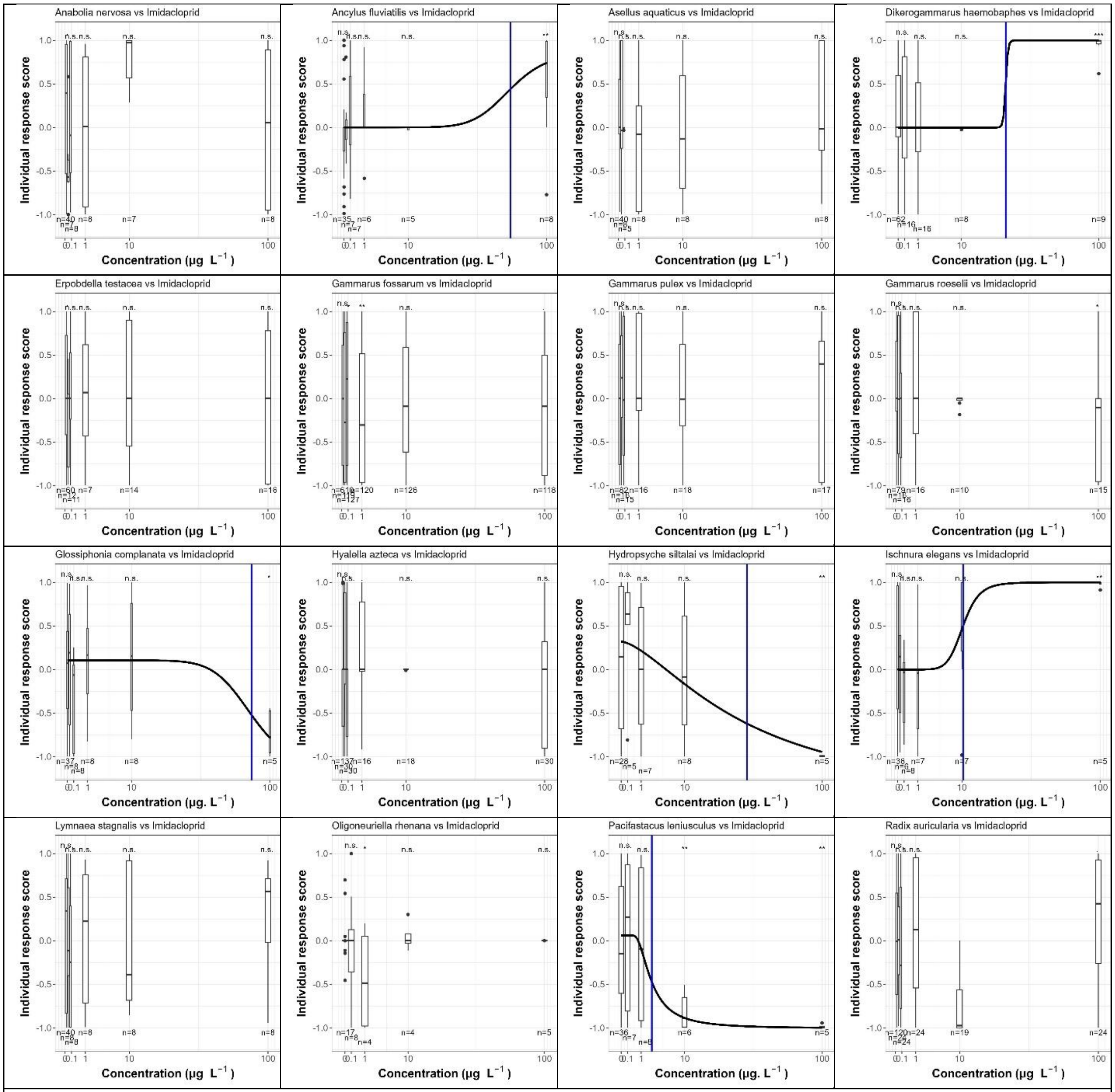

**Supplementary Figure 2:** Individual Response Score (IRS) distribution along the gradient of imidacloprid peak concentrations (freshwater species). n= number of organisms per concentration; Results of statistical significance test (difference with the pseudo-control distribution) are noticed above each condition; $EC50_{behaviour}$ is symbolized by the vertical blue line for responsive species only.

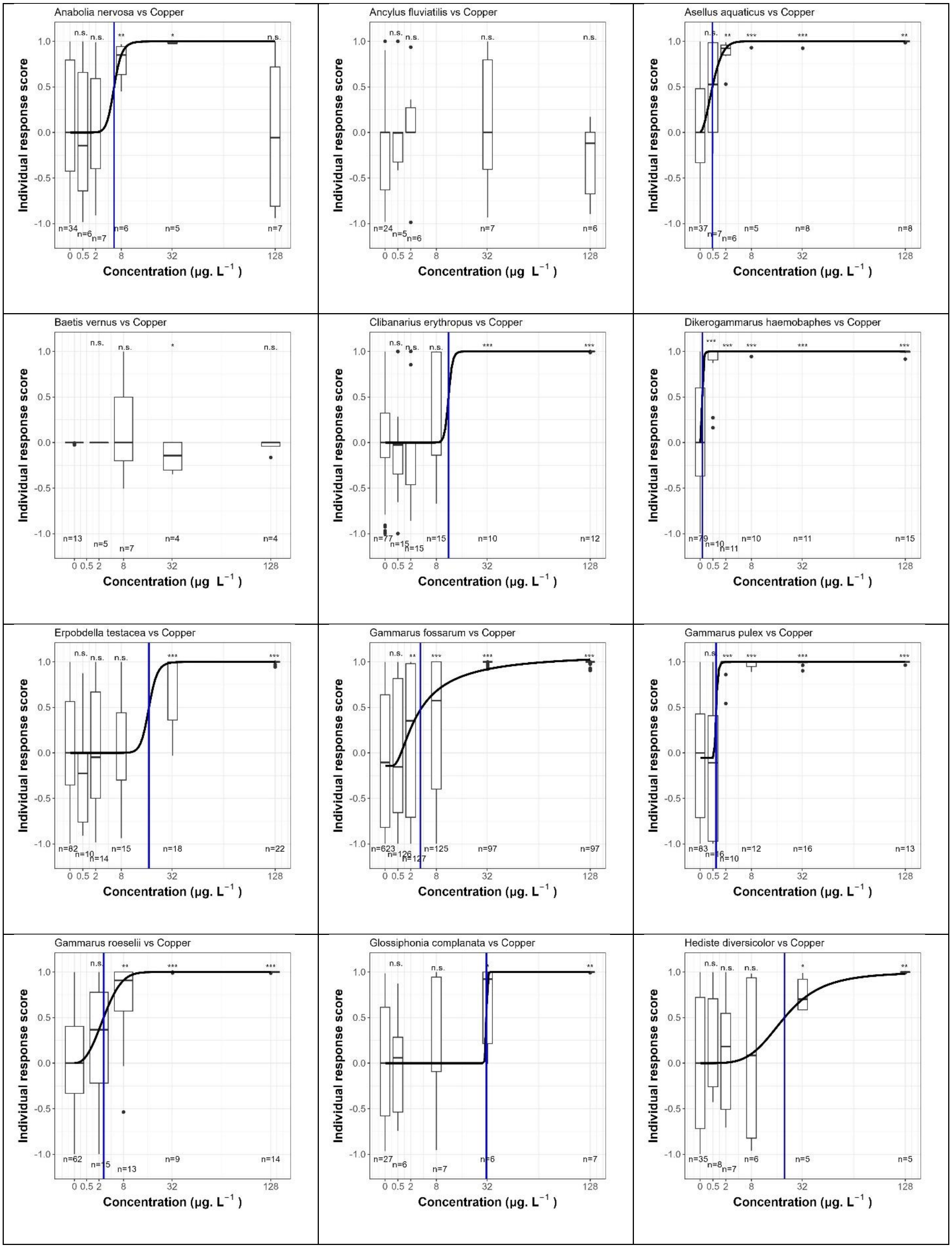
Anabolia nervosa vs Copper
Ancylus fluviatilis vs Copper
Asellus aquaticus vs Copper
Baetis vernus vs Copper
Clibanarius erythropus vs Copper
Dikerogammarus haemobaphes vs Copper
Erpobdella testacea vs Copper
Gammarus fossarum vs Copper
Gammarus pulex vs Copper
Gammarus roeselii vs Copper
Glossiphonia complanata vs Copper
Hediste diversicolor vs Copper
Individual response score
Concentration (µg. L$^{-1}$ )
0 0.5 2 8 32 128

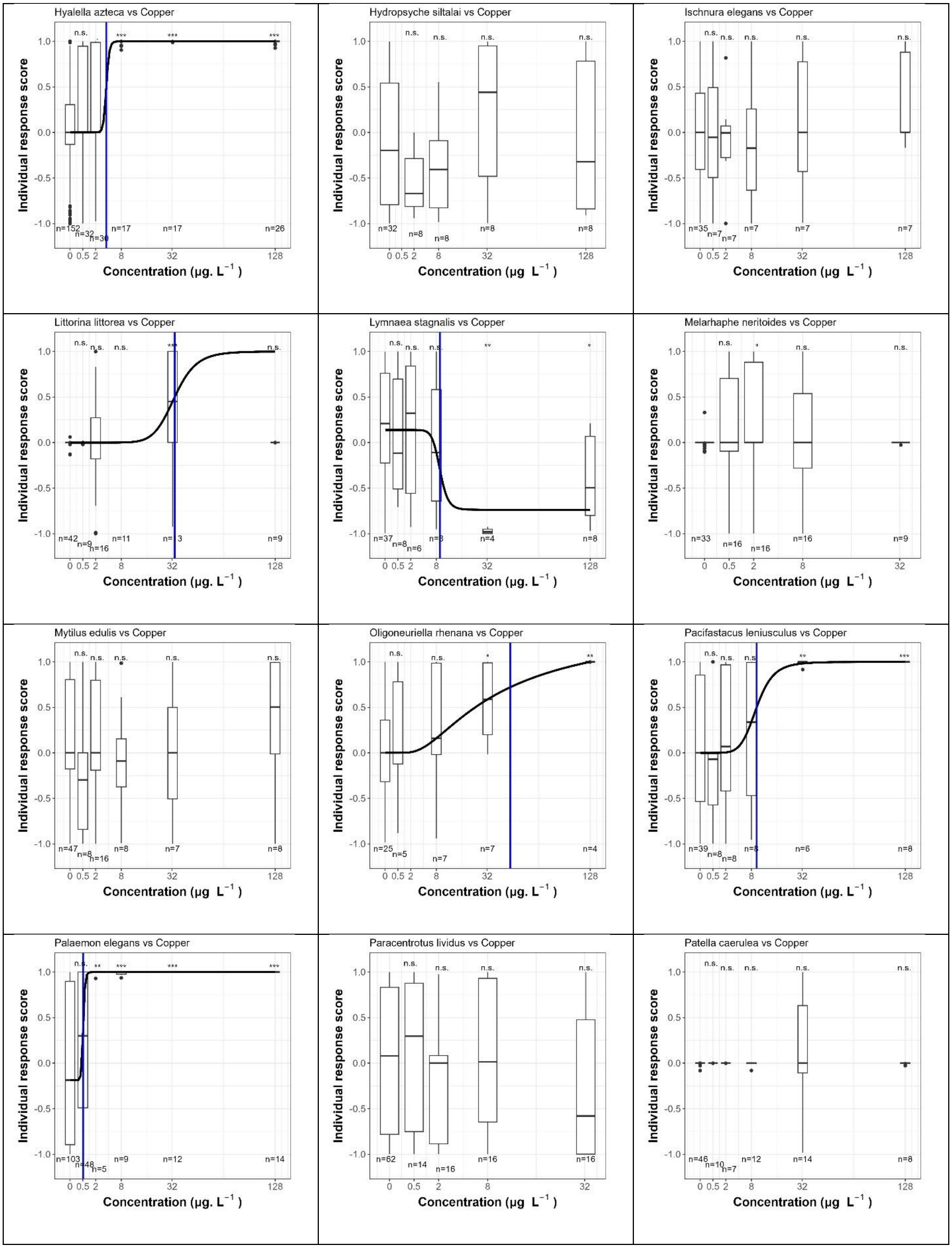
Hyalella azteca vs Copper
Hydropsyche siltalai vs Copper
Ischnura elegans vs Copper
Littorina littorea vs Copper
Lymnaea stagnalis vs Copper
Melarhaphe neritoides vs Copper
Mytilus edulis vs Copper
Oligoneuriella rhenana vs Copper
Pacifastacus leniusculus vs Copper
Palaemon elegans vs Copper
Paracentrotus lividus vs Copper
Patella caerulea vs Copper
Individual response score
Concentration (µg. L⁻¹)

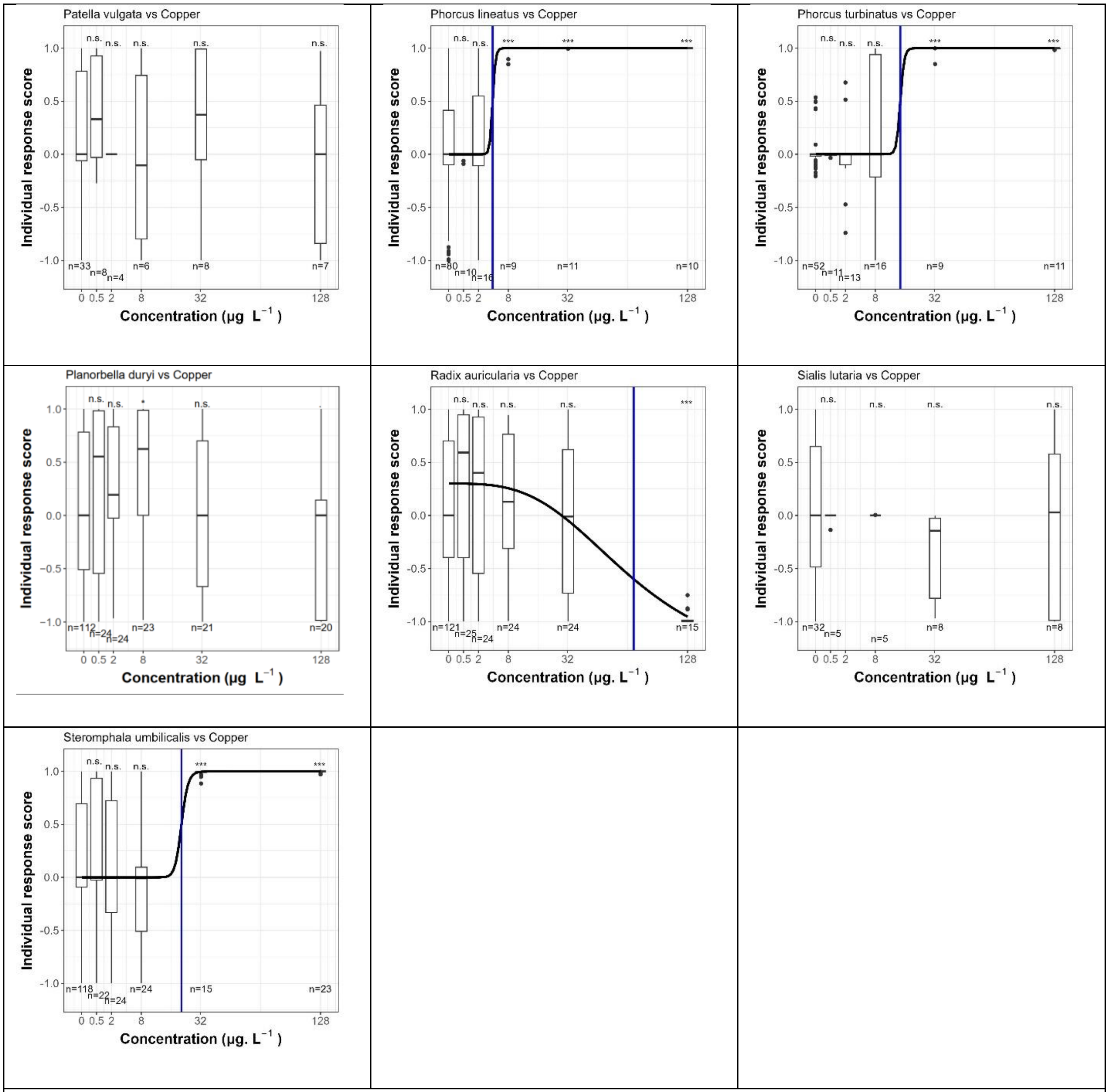

**Supplementary Figure 3:** Individual Response Score (IRS) distribution along the gradient of copper peak concentrations (freshwater and marine species). n= number of organisms per concentration; Results of statistical significance test (difference with the pseudo-control distribution) are noticed above each condition; $EC50_{behaviour}$ is symbolized by the vertical blue line for responsive species only.

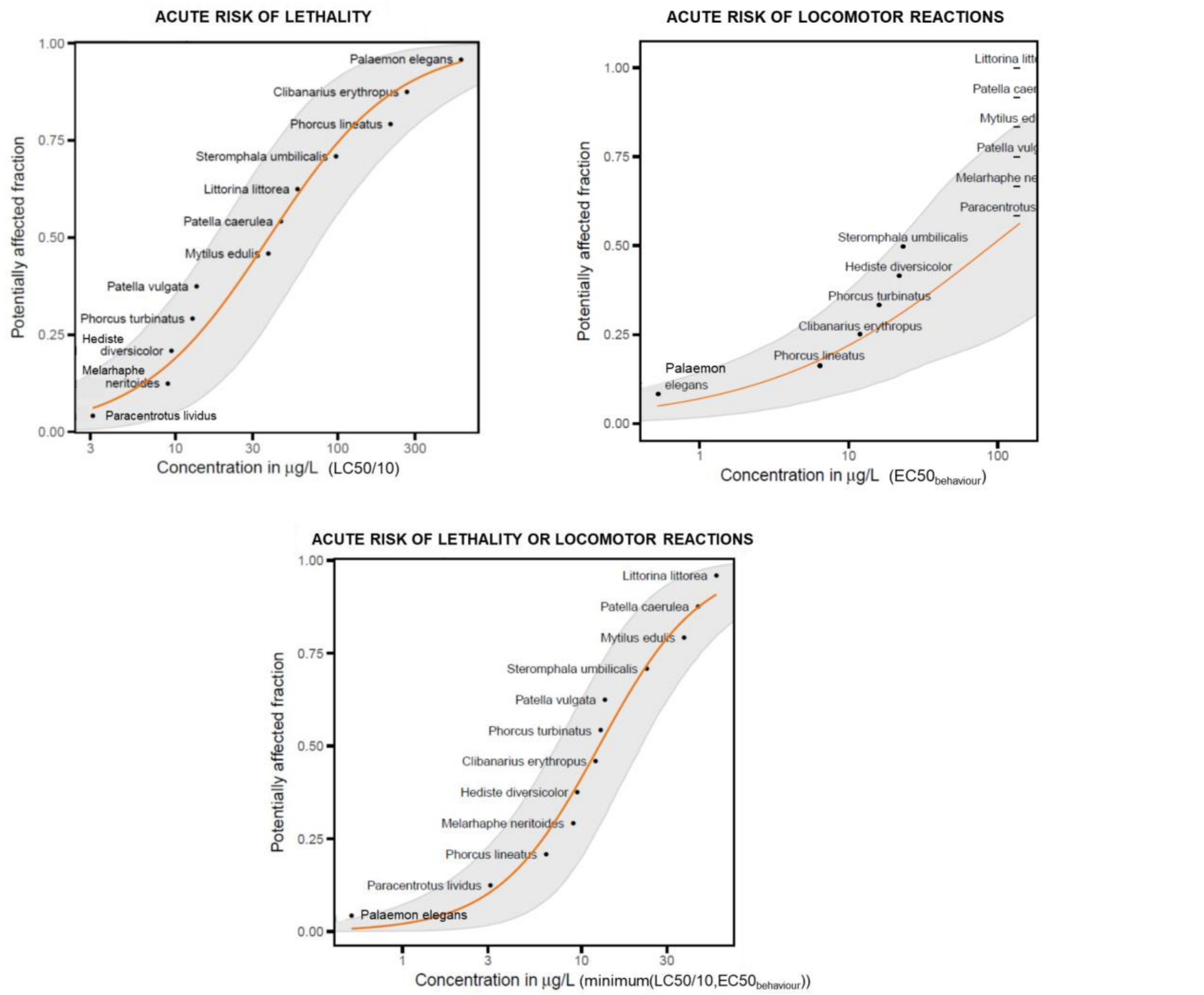

**Supplementary Figure 4:** Acute Species Sensitivity Distributions for copper in marine species established on lethality data (LC50/10), behavioural responses ($EC50_{behaviour}$) and combination of both. Log-logistic models were fitted with the MOSAIC software integrating right-censored data for $EC50_{behaviour}$ (Charles et al., 2018).

## References in Supplementary material